\documentclass[aps,twocolumn,showpacs,preprintnumbers,nofootinbib,prd,superscriptaddress,groupedaddress,10pt]{revtex4-1}
\usepackage[utf8]{inputenc}
\usepackage{graphicx,amssymb,amsmath,amsthm,amsfonts,epsfig,times,natbib}

\usepackage[usenames,dvipsnames]{color}
\usepackage{epstopdf}
\usepackage{amsmath,amssymb}
\usepackage{aas_macros}
\usepackage{tensor}
\usepackage{mathtools}
\usepackage{amsbsy}
\usepackage{bm,url}
\usepackage[linktocpage]{hyperref}
\usepackage[scaled]{beramono}
\usepackage[T1]{fontenc}

\usepackage{amsmath}
\usepackage{amsfonts}
\usepackage{amssymb}
\usepackage{bm}
\usepackage{hyperref}
\usepackage{mathrsfs}
\usepackage{graphicx}

\usepackage{ulem}
\usepackage[usenames]{color}

\newcommand{\be}{\begin{equation}}
\newcommand{\ee}{\end{equation}}
\newcommand{\ba}{\begin{eqnarray}}
\newcommand{\ea}{\end{eqnarray}}
\newcommand{\nn}{\nonumber}

\definecolor{oxfordblue}{rgb}{0.0, 0.13, 0.28}
\definecolor{burgundy}{rgb}{0.5, 0.0, 0.13}
\definecolor{darkolivegreen}{rgb}{0.33, 0.42, 0.18}
\definecolor{darkblue}{rgb}{0,0,0.5}
\definecolor{richcarmine}{rgb}{0.84, 0.0, 0.25}
\definecolor{darkblue}{rgb}{0,0,0.5}
\definecolor{venetianred}{rgb}{0.78, 0.03, 0.08}
\definecolor{skobeloff}{rgb}{0.0, 0.48, 0.45}
\hypersetup{colorlinks=true, citecolor=darkblue, linkcolor=darkblue,
urlcolor = darkblue}

\def\nn{\nonumber}

\renewcommand{\vec}[1]{\boldsymbol{#1}}

\newcommand{\ben}{\begin{enumerate}}
\newcommand{\een}{\end{enumerate}}

\def\be{\begin{equation}}
\def\ee{\end{equation}}
\def\bea{\begin{eqnarray}}
\def\eea{\end{eqnarray}}
\def\nn{\nonumber}
\newcommand{\beq}{\begin{eqnarray}}
\newcommand{\eeq}{\end{eqnarray}} 

\begin{document}
\preprint{CERN-TH-2018-184,  KCL-PH-TH/2018-34}

\title{Scattering of scalar, electromagnetic and gravitational waves from binary systems}

\author{
Lorenzo Annulli$^{1}$,
Laura Bernard$^{1}$,
Diego Blas$^{2,3}$,
Vitor Cardoso$^{1,3}$
}

\affiliation{${^1}$ Centro de Astrof\'{\i}sica e Gravita\c c\~ao  - CENTRA, Departamento de F\'{\i}sica, Instituto Superior T\'ecnico - IST, Universidade de Lisboa - UL, Av. Rovisco Pais 1, 1049-001 Lisboa, Portugal}
\affiliation{${^2}$ Theoretical Particle Physics and Cosmology Group, Department of Physics,
King’s College London, Strand, London WC2R 2LS, UK}
\affiliation{${^3}$ Theoretical Physics Department, CERN, CH-1211 Geneva 23, Switzerland}

\begin{abstract}
The direct detection of gravitational waves crowns decades of efforts in the modelling of sources and of increasing detectors' sensitivity.
With future third-generation Earth-based detectors or space-based observatories, gravitational-wave astronomy will be at its full bloom.
Previously brushed-aside questions on environmental or other systematic effects in the generation and propagation of gravitational waves
are now begging for a systematic treatment. Here, we study how electromagnetic and gravitational radiation is scattered by a binary system. Scattering cross-sections, resonances and the effect of an impinging wave on a gravitational-bound binary are worked out for the first time. The ratio between the scattered-wave amplitude and the incident wave can be of order $10^{-5}$ for known pulsars, bringing this into the realm of future gravitational-wave observatories. For currently realistic distribution of compact-object binaries, the interaction cross-section is too small to be of relevance.
\end{abstract}

\maketitle
\tableofcontents

\section{Introduction}
\subsection{Precision gravitational-wave physics}

The direct detection of gravitational waves (GWs)~\cite{Abbott:2016blz} is the first step on a long road to a new understanding of the
gravitational universe~\cite{Barack:2018yly}. Future, higher-precision observations of inspiralling black holes or neutron stars, will inform us about the number and origin of these objects, their nature and provide new information about strong-field gravity~\cite{TheLIGOScientific:2016src,Yunes:2016jcc}.
Among others, the observation of inspiralling compact objects will determine their mass and spin to levels which are all but incredible by astronomy standards~\cite{Berti:2004bd,AmaroSeoane:2007aw}; it will impose strong constraints on non-trivial radiation channels~\cite{Barausse:2016eii,Cardoso:2016olt,Arvanitaki:2016qwi,Brito:2017zvb}, and it may bring information on the local dark matter density where the process is taking place~\cite{Eda:2013gg,Barausse:2014tra,Barausse:2014pra}. Precise measurements of the gravitational waveform can tell us if the objects have nonzero tidal Love numbers,
potentially discriminating black holes from other hypothetic compact objects~\cite{Cardoso:2017cfl,Sennett:2017etc,Cardoso:2017njb,Cardoso:2017cqb}. The final, ringdown,
phase will allow us to test General Relativity~\cite{Berti:2005ys,Berti:2016lat,Yang:2017zxs}, and even to perform tests of the `black-hole' nature of the object~\cite{Cardoso:2016rao,Cardoso:2017njb,Cardoso:2017cqb}. For a review see the recent roadmap~\cite{Barack:2018yly}. 

The possibility to extend our knowledge in such fundamental questions can only be realized via {\it precision GW physics}.
This enormous potential for new science requires the careful control of any systematic factors.
Environmental effects, such as accretion disks, nearby stars, electric or magnetic fields, a cosmological constant or even dark matter, all
can possible contribute to blur what is otherwise a clear picture of compact binaries. The effects of such environment on the {\it generation}
of GWs was investigated recently~\cite{Barausse:2014tra,Barausse:2014pra}. 

\subsection{Scattering}

The effects of the environment on the {\it propagation} of GWs are usually believed to be negligible\footnote{An important counterexample are some models of dark energy. In fact, the recent observation of \cite{TheLIGOScientific:2017qsa} has been used to ruled out many candidates \cite{Ezquiaga:2017ekz,Creminelli:2017sry,Baker:2017hug}.}. If the medium is modelled as a perfect fluid,
then GWs do not couple to it and are therefore neither absorbed nor dispersed by such an environment~\cite{Grishchuk:1981fp,Deruelle:1984hq}. 
These calculations have been redone for viscous fluids, and very recently for some particle dark matter models~\cite{Baym:2017xvh,Flauger:2017ged}. See also  \cite{Dev:2016hxv,Cai:2017buj} for more promising results for dark matter models beyond the WIMP paradigm.

\begin{figure}
\begin{center}
\includegraphics[scale=0.2]{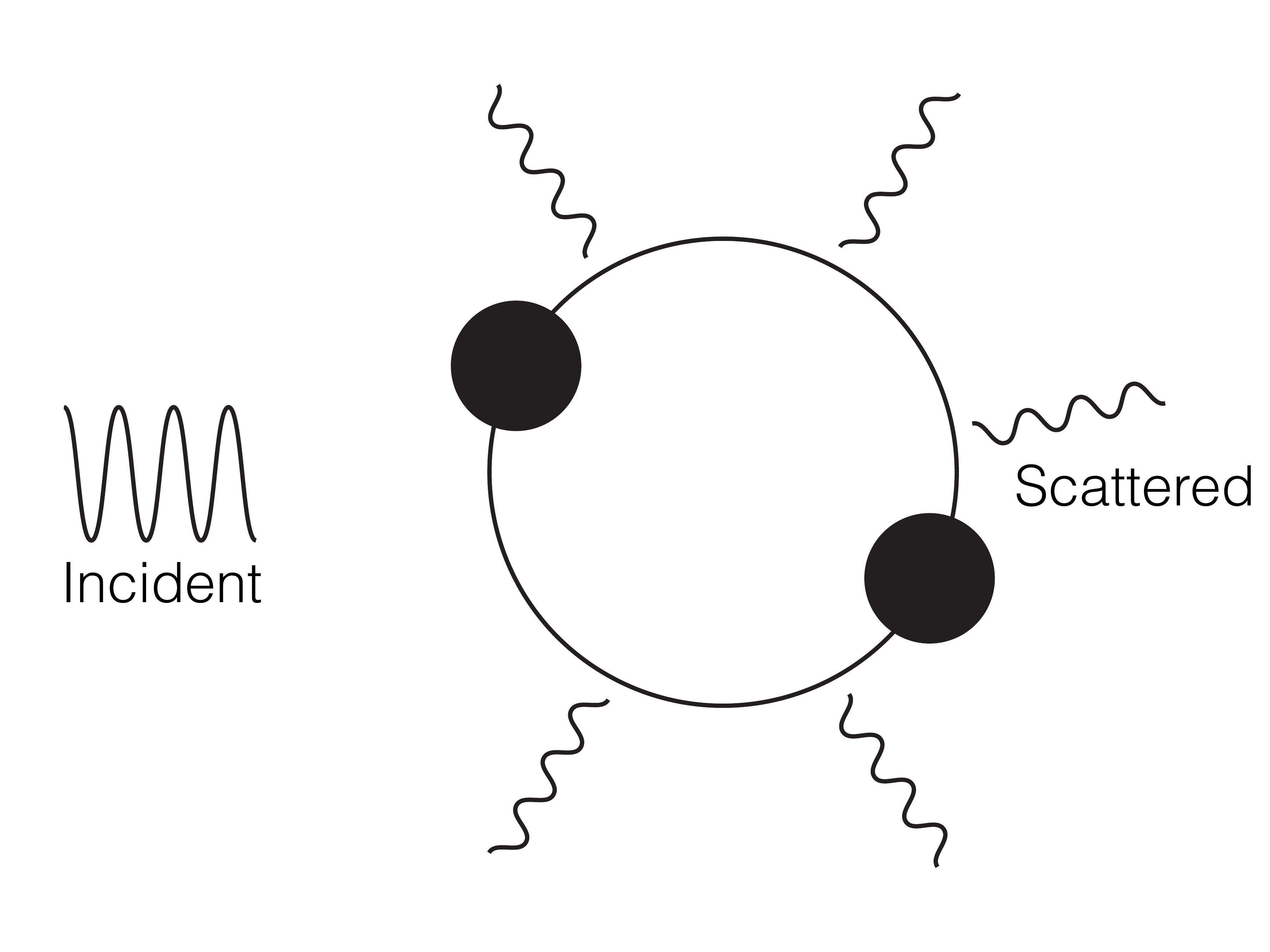}
\caption{Scattering of an incoming GW by a binary. The GW affects the motion of the binary, which in turn re-radiates and contributes to a non-trivial scattered wave.
\label{fig:scatter}}
\end{center}
\end{figure}
Here, we wish to investigate the scattering of radiation by individual obstacles, in particular a gravitationally-bound binary such as the one depicted in Fig.~\ref{fig:scatter}.
This subject remains largely unexplored, but is the gravitational counterpart of well known and observed
electromagnetic (EM) scattering phenomena (such as the Rayleigh scattering of light responsible for blue skies).
GW scattering is suppressed by the small value of the gravitational coupling constant. However, our ability to do precision measurements has increased to unforeseen levels, and will continue to do so in the next decades. 
Pulsar timing, for instance, holds the promise to overcome the smallness of this coupling and detect minor variations in the time of arrival of the radio-wave from a background of GWs~\cite{Khmelnitsky:2013lxt}. On the other hand, resonances between the impinging GWs and a binary system may enhance the effects to measurable levels.
This motivated a few studies in the past~\cite{Turner:1979yn,1978ApJ...223..285M}, both focusing on resonant interaction between a passing GW and a binary~\footnote{Our work was also motivated by a study suggesting that the modes of oscillation of stars could be excited by passing GWs~\cite{McKernan:2014hha}. The master differential equation that rules these excitations is akin to our radial displacement in the binary, due to the incoming GW. In fact, in both cases one gets resonances induced by the scattering process. Also, in the single star case there are reasons to expect that the GW signal can be the source of measurable deviations in the acoustic oscillations of the stars~\cite{Lopes:2014dba}. Moreover in the case of a binary made  of two stars, if the frequency of the excited mode is comparable with the proper orbital frequency, the scattering process can leave a signature both on the binary as a whole and on the single compact bodies in the couple. However, this topic needs further investigation to be properly clarified.}. 
An analysis for secular effects of a stochastic background of GWs was performed in \cite{Hui:2012yp}.
Here, we will take the program a step further, by computing the binary contribution to the re-radiated field, thereby truly calculating the scattered wave.

\subsection{Executive Summary}

For the sake of clarity, we outline here our main results. 
We start by working out how an incoming EM wave affects a rotating dipole. This is a classical treatment that only requires linear perturbation theory.
We use the change in the dipole moment induced by the incoming EM wave to compute the scattered radiation and the total scattering cross section. All these quantities are evaluated for an EM wave propagating along the direction of the observer and with the electric field oscillating in the plane of the orbit. 
In the high frequency limit, we recover classical results concerning scattering off oscillators. 

The equations of motion for two point-like masses on a bound orbit, \eqref{eqacceleration}, are found encapsulating the GW perturbation within a PN framework. This procedure highlights the non linear character of the Einstein equations. For GWs which are homogeneous
on length-scales larger than the characteristic orbital distance between the masses, we find the same equations of motion as those described by Turner~\cite{Turner:1979yn} and Mashhoon~\cite{1978ApJ...223..285M}. Using an angle-action formalism to treat the variation of the orbital parameters, we find that the changes in the orbital parameters are linear in the incoming GW. 
Likewise, resonances between the binary and the incoming GW happen at certain discrete GW frequencies (integer multiples of the proper orbital frequency), in agreement with previous literature~\cite{Turner:1979yn,1978ApJ...223..285M}.

We extend previous results in an important direction, by including dissipative terms and evaluating the scattered GW \eqref{hplusform}, \eqref{hcrossform} and the scattering cross section \eqref{crosssection1} for two physical configurations: (i) for GWs propagating along the direction of the angular momentum of the system (i.e. oscillating in the orbital plane), and (ii) for GWs propagating perpendicularly to the angular momentum vector (i.e. GW travelling parallel to the orbital plane). 

\subsection{Geometrical conventions}\label{Geometrical conventions}

\begin{figure}
\begin{center}
\includegraphics[scale=0.22]{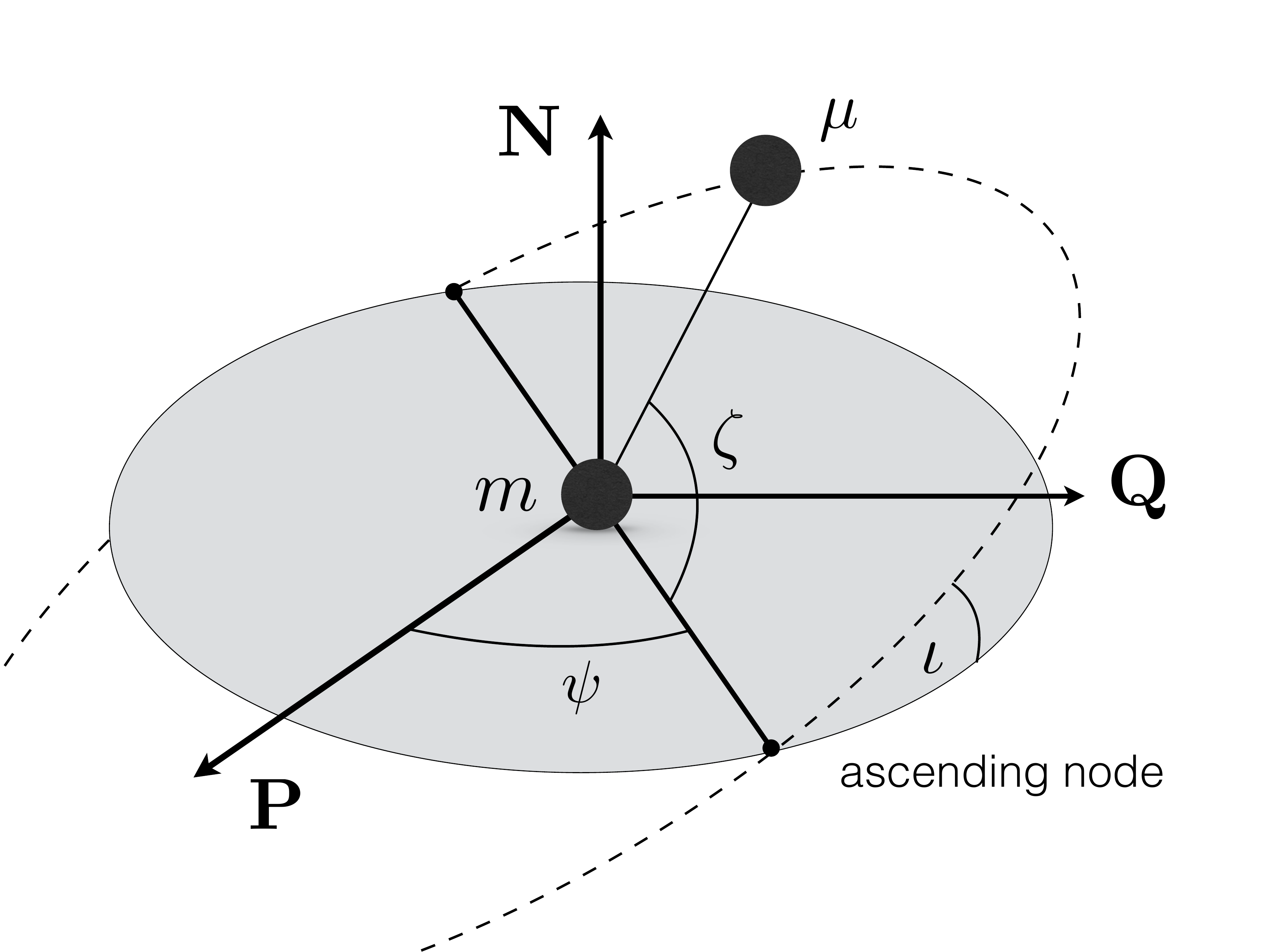}
\caption{Plane of the orbit with respect to the fixed observer basis $(\mathbf{P},\mathbf{Q},\mathbf{N})$. The angle $\zeta$ is the polar angle describing the motion of the reduced mass $\mu$ in the orbital plane, while $\mathbf{L}$ and $\iota$ are respectively the total angular momentum and the angle between this vector and the direction $\mathbf{N}$. The total mass is denoted $m$.}
\label{PlaneOfTheOrbit}
\end{center}
\end{figure}

Our calculations and description of the problem involve specific but different frames.
To avoid confusing the reader, we summarize here all the frames that we are going to use through all the paper. Consider an observer located in a direction $\mathbf{N}$, 
whose basis is $\left(\mathbf{P},\,\mathbf{Q},\,\mathbf{N}\right)$. This will be called the frame of the observer and it is fixed with respect to the observer itself. We refer the reader to Fig.~\ref{PlaneOfTheOrbit}. We will study binaries, in which the motion of the individual bodies under central forces (EM or Newtonian) are described by ellipses. We choose as unit vector $\mathbf{P}$ the one that points toward the direction of the ascending node $\mathcal{N}$. In the presence of a perturbation, this freedom to choose the ascending node no longer exists and we choose to keep the basis $\left(\mathbf{P},\,\mathbf{Q},\,\mathbf{N}\right)$ in its unperturbed configuration. Furthermore, we define $\psi$ as the angle between $\mathbf{P}$ and the ascending node $\mathcal{N}$, $\zeta$ the angle between the ascending node and the direction $\mathbf{n}$ and $\iota$ the angle between $\mathbf{N}$ and $\mathbf{L}$, where $\mathbf{L}$ is the angular momentum vector of the binary. The second frame will be the one that describes the motion of the reduced mass with respect to the center of mass. This frame is defined with respect to the following directions: $\mathbf{n}$ is the radial direction with respect to the orbital motion, $\vec{\lambda}$ is the tangent one, while $\mathbf{l}$ is directed along the angular momentum direction $\mathbf{L}$. From classical mechanics, the following relations between the binary center of mass basis and the observer basis hold~\cite{PoissonWill2014}
\begin{align}\label{basis}
\mathbf{n}  = & \left(\cos\psi\,\cos\zeta-\sin\psi\,\cos\iota\,\sin\zeta\right)\mathbf{P} \\
&  +\left(\sin\psi\,\cos\zeta+\cos\psi\,\cos\iota\,\sin\zeta\right)\mathbf{Q} +\sin\iota\,\sin\zeta\,\mathbf{N}, \notag \\
\boldsymbol{\lambda}  = & -\left(\cos\psi\,\sin\zeta+\sin\psi\,\cos\iota\,\cos\zeta\right)\mathbf{P} \\
&  +\left(\cos\psi\,\cos\iota\,\cos\zeta-\sin\psi\,\sin\zeta\right)\mathbf{Q} +\sin\iota\,\cos\zeta\,\mathbf{N}, \notag\\
\mathbf{l} = & \sin\psi\,\sin\iota\,\mathbf{P}-\cos\psi\,\sin\iota\,\mathbf{Q}  +\cos\iota\,\mathbf{N}. 
\end{align}
Note that the unperturbed case corresponds to the configuration $\psi=0$ and $\iota=\text{cst}$. In this configuration, the velocity in the center-of-mass frame is:
\begin{equation}\label{vitesse}
\mathbf{v} = \dot{r}\mathbf{n} +r\left(\dot{\zeta}+\dot{\psi}\cos\iota\right)\boldsymbol{\lambda} +r\left(\dot{\iota}\sin\zeta-\dot{\psi}\sin\iota\cos\zeta\right)\mathbf{l}\,,
\end{equation}
where $r$ is the relative position.
The frame $(\mathbf{n},\vec{\lambda},\mathbf{l})$, called CM frame in the rest of the paper, has time-varying basis with respect to the fixed observer frame. Lastly, we also introduce the proper frame of the wave $(\vec{e}_x,\vec{e}_y,\vec{e}_z)$, useful for the definition of the polarizations in both the EM and in the GR case. We denote $\alpha$ the angle between the $P$-axis and the ascending node $\mathcal{N}'$, $\beta$ the angle between the ascending node and $\mathbf{e}_{x}$ and $\kappa$ the angle between $\mathbf{e}_{z}$ and $\mathbf{N}$. We then have the following relations between the observer basis and the incoming GW basis:
\begin{align}\label{GWbasis}
\vec{e}_{x}  = & \left(\cos\alpha\,\cos\beta-\sin\alpha\,\cos\kappa\,\sin\beta\right)\mathbf{P} \\
&  \left(\sin\alpha\,\cos\beta+\cos\alpha\,\cos\kappa\,\sin\beta\right)\mathbf{Q} +\sin\kappa\,\sin\beta\,\mathbf{N}, \notag \\
\vec{e}_{y}  = & \left(\cos\alpha\,\sin\beta+\sin\alpha\,\cos\kappa\,\cos\beta\right)\mathbf{P} \\
&  \left(\sin\alpha\,\sin\beta-\cos\alpha\,\cos\kappa\,\cos\beta\right)\mathbf{Q} -\sin\kappa\,\cos\beta\,\mathbf{N}, \notag\\
\vec{e}_{z}= & -\sin\alpha\,\sin\kappa\,\mathbf{P} +\cos\alpha\,\sin\kappa\,\mathbf{Q} -\cos\kappa\,\mathbf{N}.
\end{align}
Figure \ref{PlaneOfTheOrbit} sketches the frame of the observer and of the CM.  

Finally, we will use the Keplerian parametrization of the orbit, and we perform an expansion for small eccentricities. Despite this, we will mostly concentrate on the zeroth order. Here are the parametrization we use:
\begin{align}
\label{eqEllipserho}
& r = a\left(1-e\cos(u)\right)\,, \\
\label{eqEllipsephi}
& \zeta = v\equiv 2\arctan\left[\left(\frac{1+e}{1-e}\right)^{1/2}\tan\left(\frac{u}{2}\right)\right]\,, \\
\label{eqEllipsel}
& l \equiv n\left(t-t_{0}\right) = u-e\,\sin u \,,
\end{align}
where $a$ is the semi-major axis, $e$ is the eccentricity, $u$ and $v$ are respectively the eccentric and true anomaly, $l$ is the mean anomaly, $n$ is the mean motion and $t_{0}$ is the instant of passage at the perihelion. At Newtonian order we have that $n=\omega_0$, where $\omega_0$ is the orbital frequency of the binary system.

\subsection{Acronyms and notation}

Here we summarize the recurrent acronyms that will be used in this paper:
\begin{itemize}
\item[GR] \hspace{0,1cm}    General Relativity
\item[GW] \hspace{0,1cm}    Gravitational Wave
\item[EM] \hspace{0,1cm}    Electromagnetism
\item[CM] \hspace{0,1cm}    Center of Mass
\item[SW] \hspace{0,1cm}    Scalar Wave
\item[2p] \hspace{0,1cm}    Dipole
\item[PN] \hspace{0,1cm}    Post Newtonian
\item[TT] \hspace{0,1cm}    Transverse Traceless
\item[LL] \hspace{0,1cm}    Landau Lifshitz
\item[LW] \hspace{0,1cm}    Liénard Wiechert
\end{itemize}

We will also often abbreviate the following trigonometric functions:
\begin{equation}\notag
\cos(\alpha+\beta)\equiv c_{\alpha+\beta} \text{ and }\sin(\alpha+\beta)\equiv s_{\alpha+\beta}.
\end{equation}
Furthermore, variables in bold are to be intended as vectors, while the corresponding normal ones are their corresponding magnitude. We use Greek letters to represent space-time indices and latin letters for 3-dimensional spatial indices. As the spatial indices are moved with the delta metric $\delta_{ij}$, we indifferently write them in a lower or upper position.

\section{Scattering of electromagnetic waves}\label{Scattering_EM}

We will start with an old and venerable problem, that of scattering of EM waves off obstacles~\cite{Landau:1982dva}.
This incursion will set the stage for both the scalar and gravitational case, while sharing some (many) features in common.
We want to evaluate the effect of an incoming EM wave on a binary system of two electric charges orbiting at a frequency $\omega_0$. The monochromatic EM wave propagates along the $z$ direction 
and has a frequency $\Omega$.

\subsection{Unperturbed dipole physics}\label{unpertsdipolefield}

Consider a system of two charged particles, of mass $m_1$ and $m_2$, that interact through the product between the electromagnetic potential $A^{\mu}=(\Phi/c,\vec{A})$ and four-current $J_{\mu}=(c	\rho,\vec{j})$, where $\rho$ is the charge density ($\rho=q_i \delta^3(\vec{x}-\vec{x}_i)$) and $\vec{j}=\rho \vec{v}$ is the current density.
We take these charges to interact only through the Coulomb force in Minkowski flat spacetime with metric $\eta_{\mu\nu}=\mathrm{diag}(-1,1,1,1)$. Using $x^{\mu}=(x^0,x^1,x^2,x^3)=(c t,x,y,z)$ as coordinates, where $c$ is the speed of light in vacuum, the action that describes this system is
\beq
S &=&\int \mathrm{d}^3x \mathrm{d}t \left[-\frac{F_{\mu\nu}F^{\mu\nu}}{4\mu_0}-A_{\mu}^1 J_{\mu}^2-A_{\mu}^2 J_{\mu}^1\right]\nonumber\\
&&  -c^2\int \mathrm{d}\tau \left(m_1+m_2\right)\,,\label{action1}
\eeq
in which $\mu_0$ is the magnetic vacuum permeability, $F_{\mu\nu}$ the antisymmetric electromagnetic tensor defined as $F_{\mu\nu}=\partial_{\mu}A_{\nu}-\partial_{\nu}A_{\mu}$ and $\mathrm{d}\tau=\mathrm{d}t\sqrt{1-v^2/c^2}$, where $v^2$ is the square of the three-velocity $v^i=\mathrm{d}x^i/\mathrm{d}t$. From now on, we restrict ourselves to the small velocities case, dropping all the special-relativistic terms. With all these assumptions, the $i-th$ component of the equations of motion for each particles is
\begin{align}\label{eoMunperturbed}
m_1\ddot{\vec{r}}_1^i=\frac{q_1 q_2 (\vec{r}_1-\vec{r}_2)^i}{ \lvert \vec{r}_1-\vec{r}_2 \lvert^3}, \;\;
m_2\ddot{\vec{r}}_2^i=\frac{q_1 q_2 (\vec{r}_2-\vec{r}_1)^i}{ \lvert  \vec{r}_2-\vec{r}_1 \lvert^3} \,,
\end{align}
where $i=(1,2,3)$, $\vec{r}_{1(2)}$ represents the position vector of particle $1(2)$, $q_1,q_2$ are the electric charges and the double dot sign means a second derivative with respect to time $t$. In the center-of-mass frame, the center of mass vector position has zero second time derivative ($\ddot{\vec{R}}_{\rm CM}=0$), while, defining the relative position vector with respect to the radial direction defined in Sec.~\ref{Geometrical conventions} as $ \vec{r}\equiv \vec{r}_1-\vec{r}_2 =r \vec{n}$, the equations for the relative motion become
\begin{equation}
\ddot{\vec{r}}=\frac{1}{\mu}\frac{q_1 q_2}{ \mid \vec{r}\lvert^2}\vec{n},
\end{equation}
where $\mu$ is the reduced mass of the system,
\be
\mu=\frac{m_1 m_2}{m_1+m_2}\,.
\ee
We define the total mass as $m=m_1+m_2$.
Since the Coulomb force is central, the total angular momentum of the system is conserved and the motion happens on a fixed plane. The solution to the equations of motion, in analogy with the Newtonian ones, have the characteristic shape of a conic section, depending on the energy of the particles. Since we are interested in bound systems, we assume that the energy will be the one associated with bound orbits. 

We focus on the case in which the dipole is composed of two particles with equal and opposite charge and equal mass:
\beq
-q_2&=&q_1=q\,,\nonumber\\
m_1&=&m_2=M\,.\label{charge_def}
\eeq
From the equations \eqref{eoMunperturbed}, we find that the center of mass is fixed, the angular momentum of the system is constant and the motion lies in the orbital plane. The orbit of the binary can be directly obtained from
\begin{equation}
 \ddot{r}-\frac{4 L^2}{M^2 r^3}=-\frac{2q^2}{M r^2}\,,
 \end{equation}
where $L=M r^2\dot{\phi}/2$ is the magnitude of the angular momentum vector of the system and $\phi$ is the angle describing the motion of the reduced mass in the plane of the orbit (polar angle).
Defining the dipole vector ($\vec{d}$) as
\begin{equation}
\vec{d}=q_1\vec{r}_1+q_2\vec{r}_2=\mu \left(\frac{q_1}{m_1}-\frac{q_2}{m_2}\right)\vec{r}\,,
\end{equation}
where $\vec{r}$ is the proper radius of the system (relative position vector in the dipole case). Introducing the vector between the CM and the observer, of magnitude $R_0$ and unit direction $\hat{\vec{R}}_0$, the generated EM wave has vector potential, electric field and magnetic one given by
\begin{equation}\label{EandMfields}
\vec{A}=\frac{1}{c R_0}\dot{\vec{d}},\,\vec{H}=\frac{1}{c^2 R_0}\ddot{\vec{d}}\times\hat{\vec{R}}_0,\, \vec{E}=\frac{1}{c^2 R_0}(\ddot{\vec{d}}\times\hat{\vec{R}}_0)\times\hat{\vec{R}}_0.
\end{equation}
This is a well-known result, a dipole emits only if it is accelerated. Finally, the expression for the intensity of the emitted energy is given by \cite{Landau:1982dva}:
\begin{equation}\label{genEmittDipole}
\mathrm{d}I=c\frac{H^2}{4\pi}R_0^2 \mathrm{d}o\rightarrow I=\frac{2}{3c^3}\ddot{d}^2,
\end{equation}
where we averaged over one period of the orbit and $\mathrm{d}o$ is the solid angle in the $\hat{\vec{R}}_0$ direction.

\subsection{Scattering from a rotating dipole}
\subsubsection{Initial considerations}

The binary above is now hit by an EM wave described by a vector potential $A^{\mu}_{\Omega}$.
For definiteness, the wave propagates along the $\vec{e}_z$ axis, parallel to the direction of the observer $\mathbf{N}$ and to the angular momentum of the system $\vec{L}$. In this way, the $x-y$ plane of the orbital frame, of the observer and also of the wave are all parallel between each other and perpendicular to the $z$ direction of the observer. 

The action~\eqref{action1} needs to be complemented by adding both the scalar and the vector potentials of the perturbation,
\begin{align}
A^{\mu}_1 \rightarrow A^{\mu}_1+A^{\mu}_{\Omega}=\left(\frac{\Phi_1}{c}+\frac{\Phi_{\Omega}}{c},\vec{A}_1+\vec{A}_{\Omega}\right),\\
A^{\mu}_2 \rightarrow A^{\mu}_2+A^{\mu}_{\Omega}=\left(\frac{\Phi_2}{c}+\frac{\Phi_{\Omega}}{c},\vec{A}_2+\vec{A}_{\Omega}\right)\,.
\end{align}
Using the definitions of EM fields~\footnote{In order to pass to the old vectorial picture, in this section $\vec{B}=\mu_0 \vec{H}$ is the magnetic field in vacuum.} and potentials,
\begin{equation}
\vec{E}=-\vec{\nabla}\Phi -\frac{\partial}{\partial t}\vec{A}\;\;\; \text{ and }\;\;\; \vec{B}=\vec{\nabla}\times \vec{A}\,,
\end{equation}
one finds,
\begin{subequations}\label{AppForEoMinEBecc}
\begin{align}
m_1 \vec{a}_1=&q_1 \vec{E}_2+q_1 \vec{E}_{\Omega}+q_1 \vec{v}_1\times \vec{B}_2+q_1 \vec{v}_1\times \vec{B}_{\Omega},\\
m_2 \vec{a}_2=&q_2 \vec{E}_1+q_2 \vec{E}_{\Omega}+q_2 \vec{v}_2\times \vec{B}_1+q_2 \vec{v}_2\times \vec{B}_{\Omega}.
\end{align}
\end{subequations}
Dropping the last two terms of Eqs.~\eqref{AppForEoMinEBecc} by assumptions of small internal velocities compared to the speed of light, we get
\begin{subequations}\label{EoMinEBecc}
\begin{align}
m_1 \vec{a}_1=&q_1 \vec{E}_2+q_1 \vec{E}_{\Omega},\\
m_2 \vec{a}_2=&q_2 \vec{E}_1+q_2 \vec{E}_{\Omega}.
\end{align} 
\end{subequations}
Finally, in the CM frame we have
\begin{align}
\ddot{\vec{R}}_{\rm CM}&=\frac{q_1+q_2}{m_1+m_2}(\vec{E}_{\Omega})_{\rm CM},\\
\label{CMacc}
\ddot{\vec{r}}&=\frac{1}{\mu}\frac{q_1 q_2}{ \lvert r\lvert^2}\vec{n}+\left(\frac{q_1}{m_1}-\frac{q_2}{m_2}\right)\left(\vec{E}_{\Omega}\right)_{\rm CM}.
\end{align}
where $\left(\vec{E}_{\Omega}\right)_{\rm CM}$, means that the quantity under consideration has to be properly expressed in the CM frame. Using the equations of motion~\eqref{EoMinEBecc} and transforming all the quantities in the CM frame, we find the total angular momentum variation in time,
\begin{equation}\label{L_TimeDerivative}
\frac{\mathrm{d} \vec{L}}{\mathrm{d} t}=
\mu \frac{2q}{M} \vec{r}\times\left (\vec{E}_{\Omega}\right)_{\rm CM}\,.
\end{equation}
Here, we used already the specific setup described by Eq.~\eqref{charge_def}.

\subsubsection{Equations of motion}

As we have shown in Eq.~\eqref{L_TimeDerivative}, the time variation of the angular momentum is given by the cross product of the relative position vector and the external perturbing force $\vec{F}_{\Omega}$,
\begin{equation}
\dot{\vec{L}}\sim\vec{r}_{12}\times \vec{F}_{\Omega}\,.
\end{equation}
An electric field on the plane of the orbit changes the magnitude of the angular momentum, but not its direction. We should highlight that this simplification still captures the dynamics of the scattering, allowing us to give an analytic treatment of the process. In order to further simplify our calculations, we consider the unperturbed motion happening in circular orbits. Therefore, the equations that describe the perturbation of such kind of trajectory are given by
\begin{subequations}\label{system1}
\begin{align}
\label{system1rad}
 \ddot{r}-r\dot{\phi}^2&=-\frac{2 q^2}{M r^2}+\frac{ q E_{\Omega}}{ M}\left(c_{\gamma-\Omega t-\phi(t)}+c_{\gamma+\Omega t-\phi(t)}\right),\\
\label{system1tan}
 2 \dot{r}\dot{\phi}+r \ddot{\phi}&=\frac{q E_{\Omega}}{ M} \left(s_{\gamma-\Omega t-\phi(t)}+s_{\gamma+\Omega t-\phi(t)}\right),
\end{align}
\end{subequations}
where $\gamma$ is the angle between the direction of polarization of the electric field and the $P$ direction, in the plane of the orbit. 
Here, and in this section only, $\phi$ is the polar angle describing the orbital motion in the $x-y$ plane. The constant $E_\Omega$ is the amplitude of the electric field.
The presence of the perturbation in the right hand side of the second equation spoils the constancy of the angular momentum but, at first order in $E_{\Omega}$, one can find the relation between $\dot{\phi}$ and $L$. Let us write,
\beq
2 \dot{r}\dot{\phi}+r \ddot{\phi}&\equiv&\frac{2}{M}\frac{1}{r}\frac{\mathrm{d}}{\mathrm{d} t}\left(\frac{M}{2}r^2\dot{\phi}\right)=\frac{2}{M}\frac{1}{r}\frac{\mathrm{d}}{\mathrm{d} t}L(t),\label{Kepler_system}
\eeq
where $L(t)$ is the angular momentum magnitude. Since without any external perturbation the angular momentum is conserved (and equal to a constant $L_{\circ}$), we can expand $L(t)$ in powers of the electric field
\begin{equation}
L(t)=L_{\circ}+E_{\Omega} L_1(t)+{\cal O}\left(E_{\Omega}^2\right)\,.
\end{equation} 
Making use of this, a similar expansion for $r(t)$ and for $\phi(t)$ can be found,
\begin{equation}\label{power in r}
r(t)=r_{\circ}+E_{\Omega} g(t)+{\cal O}\left(E_{\Omega}^2\right)\,,
\end{equation}
\begin{equation}\label{power in phi}
\phi(t)=\phi(0)+t \, \dot{\phi}=\phi(0)+t(\omega_0+E_{\Omega} Z_p+{\cal O}\left(E_{\Omega}^2\right))\,,
\end{equation}
where $r_{\circ}$ is the orbital radius of the unperturbed motion, $\phi(0)=\phi_0$ is the initial angular position of the reduced mass in the $x-y$ plane and $Z_p$ is the first order correction in the orbital frequency due to the external perturbation. Using Eqs.~\eqref{system1} we find
\begin{equation}\label{Ldot}
\dot{L}_1(t)=\frac{q\, \, r_{\circ}}{2} \left(s_{\gamma_0-\phi_0-t \Omega - t\omega_0}+s_{\gamma_0-\phi_0+t \Omega - t\omega_0}\right)\,.
\end{equation}
where we kept only the zero order in the $\phi(t)$ expansion because $\dot{L}_1$ is already a first order quantity. For the unperturbed circular motion, $\dot{\phi}=\omega_0$ is constant. Thus, integrating Eq.~\eqref{Ldot} with $\phi(t)=\omega_0 t$, one finds,
\beq
L_1(t)&=&\int_0^t \mathrm{d}t'[\dot{L}_1(t')]\nonumber\\
&=&q r_{\circ} \left(\frac{\omega_0 c_{\gamma-\phi_0}}{\Omega ^2-\omega_0^2}+\frac{c_{\gamma-\phi_0+t \Omega - t\omega_0}}{2 (\omega_0-\Omega )}+\frac{c_{\gamma-\phi_0-t \Omega - t\omega_0}}{2 (\omega_0+\Omega)}\right)\,.\nonumber
\eeq
Finally, the total angular momentum to first order in the external field is
\beq
L(t)&=&L_{\circ}+E_{\Omega} L_1(t)\nonumber\\
&=& L_{\circ}+\frac{E_{\Omega} q r_{\circ}\omega_0 c_{\gamma-\phi_0}}{\Omega^2-\omega_0^2}\nonumber\\
&& \hspace{-.3cm}+\frac{E_{\Omega} q r_{\circ}}{2}\bigg[\frac{c_{\gamma-\phi_0+t \Omega - t\omega_0}}{\omega_0-\Omega}+\frac{c_{\gamma-\phi_0-t \Omega - t\omega_0}}{\omega_0+\Omega}\bigg].\label{L_dot}
\eeq
From the definition of angular momentum, from Eqs.~\eqref{power in r} and \eqref{power in phi} and  up to ${\cal O}\left(E_{\Omega}^2\right)$,
\beq
L(t)&=&\frac{1}{2}M r(t)^2 \dot{\phi}(t)\nonumber\\
&=& \frac{M}{2} r_{\circ}^2\omega_0+\left(\frac{M}{2}r_{\circ}(r_{\circ}Z_p+2\omega_0 g(t))\right)E_\Omega\,.
\eeq
We can compare with Eq.~\eqref{L_dot} order by order, to get
\beq
L_{\circ}=\frac{M}{2} r_{\circ}^2\omega_0\,,
\eeq	
at order zero, and
\beq
Z_p&=&\frac{q}{M r_{\circ}}\left(\frac{2\omega_0 c_{\gamma-\phi_0}}{\Omega ^2-\omega_0^2}+\frac{c_{\gamma-\phi_0+t \Omega - t\omega_0}}{(\omega_0-\Omega )}+\frac{c_{\gamma-\phi_0-t \Omega - t\omega_0}}{(\omega_0+\Omega)}\right)\nonumber\\
&&-\frac{2 \omega_0 g(t)}{r_{\circ}}\,.
\eeq
Now that we have used the Keplerian polar equation to get the angular perturbation due to the incoming wave, we substitute this result in $\dot{\phi}^2$ in the radial equation \eqref{system1rad} in order to find the equation governing $g(t)$. Then substituting the expansions  given by Eqs.~\eqref{power in r}-~\eqref{power in phi},
\beq
\ddot{g}(t)&-&r_\circ(\omega_0+E_\Omega Z_p)^2=-\frac{2q^2}{M(r_\circ+E_\Omega g(t))}\nonumber\\
&+&\frac{ q E_{\Omega}}{ M}\left(c_{\gamma-\Omega t-\phi_0-\omega_0 t}+c_{\gamma+\Omega t-\phi_0-\omega_0 t}\right)\,,\label{g_of_t_equation}
\eeq
we get, at zero order in $E_\Omega$, the relation between the Newtonian orbital frequency and the characteristics of the binary,
\beq
\omega_0^2=\frac{2 q^2}{M r_{\circ}^3}\,.\label{omega_zero}
\eeq
Substituting $M$ obtained by the equation above in the first order expansion of Eq.~\eqref{g_of_t_equation}, we find a differential equation for $g(t)$.
\begin{widetext}
\beq
\ddot{g}(t)+\omega_0^2 g(t)+\frac{2r_{\circ}^3 \omega_0^4  c_{\gamma-\phi_0}}{q(\omega_0^2-\Omega^2)}=\frac{r_{\circ}^3 \omega_0^2 (\Omega-3\omega_0)c_{\gamma-\phi_0+t \Omega - t\omega_0}}{2q(\Omega-\omega_0)}+\frac{r_{\circ}^3 \omega_0^2 (\Omega+3\omega_0)c_{\gamma-\phi_0-t \Omega - t\omega_0}}{2q(\Omega+\omega_0)}
\,,\label{perturbationLaprroxint}
\eeq
The equation above represents a driven harmonic oscillator with multiple resonant frequencies, whose solution is given by
\beq
g(t)&=&k_1 \cos(t\omega_0)+k_2 \sin(t\omega_0)+\frac{r_\circ^3}{q}\left[\frac{2\omega_0^2  c_{\gamma-\phi_0}}{\Omega^2-\omega_0^2}-\frac{\omega_0^2 (\Omega-3\omega_0)c_{\gamma-\phi_0+t \Omega -t\omega_0}}{2\Omega(\Omega^2-3\Omega\omega_0+2\omega_0^2)}-\frac{\omega_0^2 (\Omega+3\omega_0)c_{\gamma-\phi_0-t \Omega -t\omega_0}}{2\Omega(\Omega^2+3\Omega\omega_0+2\omega_0^2)}\right].\label{gOftcircular}
\eeq
\end{widetext}
in which $k_1$ and $k_2$ are integration constants. We set in the following the two constants of integration to zero. Finally, we can evaluate $Z_p$ considering the explicit solution for $g(t)$ given by Eq.~\eqref{gOftcircular} with $k_1=k_2=0$,
\beq
Z_p&=&\frac{r_\circ^2}{q}\bigg[\frac{3\omega_0^3 c_{\gamma-\phi_0}}{\omega_0^2-\Omega^2}-\frac{\omega_0^2 (\Omega^2-4\Omega\omega_0+6\omega_0^2)c_{\gamma-\phi_0+t\Omega-t\omega_0}}{2\Omega(\Omega-2\omega_0)(\Omega-\omega_0)}\nonumber\\
&+&\frac{\omega_0^2 (\Omega^2+4\Omega\omega_0+6\omega_0^2)c_{\gamma-\phi_0-t\Omega-t\omega_0}}{2\Omega(\Omega+2\omega_0)(\Omega+\omega_0)}
\bigg]\,.\label{Zp}
\eeq
The roots of the denominators in the solution for $g(t)$ are
\beq
\bigg\{-2\omega_0,-\omega_0,0,\omega_0,2\omega_0\bigg\}\,.\nonumber
\eeq
The negative values are solution because of the symmetry of the problem, but they are not adding any physics to the positive ones, so we will consider only $0,\omega_0,2\omega_0$. Let's evaluate the limit of $r(t)$ for these roots,
\begin{subequations}
\beq
\lim_{\Omega\rightarrow 0}r(t)&=&r_{\circ}-\frac{2E_\Omega r_\circ^3 c_{\gamma-\phi_0}}{q}+\frac{E_\Omega r_\circ^3}{q}\bigg(\frac{7}{4}c_{\gamma-\phi_0-t \omega_0}\bigg)\nonumber\\
&-&\frac{E_\Omega r_\circ^3}{q}\bigg(\frac{3}{2}t \, \omega_0 s_{\gamma-\phi_0-t \omega_0}\bigg)\label{Omega_0limit}\,,
\eeq
\beq
\lim_{\Omega\rightarrow \omega_0}r(t)&=&r_{\circ}-\frac{E_\Omega r_\circ^3}{q}\bigg(\frac{1}{3}c_{\gamma-\phi_0- 2 t \omega_0}\bigg)\nonumber\\
&+&\frac{E_\Omega r_\circ^3}{q}\bigg(t\,\omega_0 s_{\gamma-\phi_0}\bigg)\,,
\eeq
\beq
\lim_{\Omega\rightarrow 2\omega_0}r(t)=\infty\,.
\eeq
\end{subequations}
In the high-frequency limit, the reasoning described before does not hold because the effect of the external field lives on a timescale much shorter than the one associated with the proper rotation of the binary, such that we can neglect the free motion of the system during one (or few) period of oscillation of the external electric field. So, we can just consider that 
\beq
\lim_{\Omega\rightarrow \infty}r(t)=r_{\circ}\,.
\eeq
From these results we see that resonant phenomena appear depending on the ratio between the incoming and the orbital frequency. Especially, in the $\Omega=0$ limit the radial motion of the reduced mass has a secular instability given by the last term of Eq.~\eqref{Omega_0limit}. This term can be understood thinking that the low frequency limit of our scattering corresponds to a perfect dipole inside a capacitor: in the large time limit, the two particles are dragged away from each other. In the $\Omega=\omega_0$ case there is also such a secular term, but it can be set to zero with an appropriate choice of the initial condition. Finally, the $\Omega=2\omega_0$ case corresponds to a proper resonance, meaning that the amplitude of the motion for that value is infinite.

\subsubsection{Scattered Fields}

Having solved the perturbed equations of motion, we can find the scattered electrical field, energy and the total cross section. 
When the system interacts with the external perturbation, the total field will contain a perturbed dipole term. In addition, the CM may contribute to the scattered field; we denote this contribution $\vec{E}_{\rm LW}$, where LW stands for Li\'enard–Wiechert,
\begin{equation}
\vec{E}_{\rm scattered}=\vec{E}_{2p}+\vec{E}_{\rm LW}\,.
\end{equation}
We express our results in the fixed observer frame, using
\beq
\hat{\vec{R}}_0&=&\cos\delta\cos\xi\,{\mathbf{P}}+\sin\delta\cos\xi\,{\mathbf{Q}}+\sin\xi\,{\mathbf{N}}\,,\\
\vec{n}(t)&=&\cos{\phi(t)}\,{\mathbf{P}}+\sin\phi(t)\,{\mathbf{Q}}\,,\label{n_of_t}\\
\vec{\mathcal{E}}_{\Omega}(t)&=&\cos\gamma\,{\mathbf{P}}+\sin\gamma\,{\mathbf{Q}}\,,
\eeq
where $\delta$ and $\xi$ are the angles that characterize the position of the unitary vector $\hat{\vec{R}}_0$ with respect to the CM, in the fixed observer frame; $\phi(t)$ is given by Eqs.~\eqref{power in phi} and \eqref{Zp}, $\gamma$ is the direction of the linear polarization of the electric field in the orbital plane and $\vec{\mathcal{E}}_\Omega$ is the unitary vector in the direction of the external electric field ($\vec{E}_\Omega=E_\Omega \vec{\mathcal{E}}_\Omega$). Since we are considering the motion of a dipole in which the total charge is zero, the contribution from the CM acceleration is zero, as shown in appendix~\ref{Radiation from the CM}.

The vector potential has contributions from the unperturbed dipole and a contribution from the perturbed part, induced by the incoming EM wave. Particularly, 
getting the electric field $\vec{E}$ from the vector potential $\vec{A}$, we find the same functional expression~\eqref{EandMfields}, but containing the acceleration of the dipole given by Eq.~\eqref{CMacc}. Therefore, using the definitions of dipole fields in Eq.~\eqref{EandMfields} and the expression for the radial separation \eqref{gOftcircular} we find,
\begin{widetext}
\beq
\vec{E}_{2p}(t)&=&-\frac{ q r_{\circ}\omega_0^2}{c^2 R_0}\left[\left(\vec{n}(t)\times\hat{\vec{R}}_0\right)\times\hat{\vec{R}}_0\right]+\frac{E_{\Omega} r_{\circ}^3 \omega_0^2 c_{\Omega t}}{c^2 R_0}\left[ \left( \vec{\mathcal{E}}_{\Omega}\times\hat{\vec{R}}_0\right) \times \hat{\vec{R}}_0\right]+\frac{4 E_{\Omega}  \omega_0^4 r_{\circ}^3 c_{\gamma-\phi_0}}{c^2 R_0 (\Omega^2-\omega_0^2)}\left[\left(\vec{n}(t)\times\hat{\vec{R}}_0\right)\times\hat{\vec{R}}_0\right]\nonumber\\
&-&\frac{2 E_{\Omega} r_\circ^3 \omega_0^4}{c^2 R_0}\bigg[\frac{ (\Omega+3\omega_0)c_{\gamma-\phi_0-t\Omega-t\omega_0}}{2 \Omega (\Omega^2+3 \Omega \omega_0+2 \omega_0^2)}+\frac{(\Omega-3\omega_0)c_{\gamma-\phi_0+t\Omega-t\omega_0}}{2 \Omega (\Omega^2-3 \Omega \omega_0 +2\omega_0^2)} \bigg]\left[\left(\vec{n}(t)\times\hat{\vec{R}}_0\right)\times\hat{\vec{R}}_0\right] \,.\label{E2pGeneral}
\eeq
\beq
\vec{H}_{2p}(t)&=&-\frac{ q r_{\circ}\omega_0^2}{c^2 R_0}\left[\vec{n}(t)\times\hat{\vec{R}}_0\right]+\frac{E_{\Omega} r_{\circ}^3 \omega_0^2 c_{\Omega t}}{c^2 R_0}\left[ \vec{\mathcal{E}}_{\Omega}\times\hat{\vec{R}}_0\right]+\frac{4 E_{\Omega}  \omega_0^4 r_{\circ}^3 c_{\gamma-\phi_0}}{c^2 R_0 (\Omega^2-\omega_0^2)}\left[\vec{n}(t)\times\hat{\vec{R}}_0\right]\nonumber\\
&-&\frac{2 E_{\Omega} r_\circ^3 \omega_0^4}{c^2 R_0}\bigg[\frac{ (\Omega+3\omega_0)c_{\gamma-\phi_0-t\Omega-t\omega_0}}{2 \Omega (\Omega^2+3 \Omega \omega_0+2 \omega_0^2)}+\frac{(\Omega-3\omega_0)c_{\gamma-\phi_0+t\Omega-t\omega_0}}{2 \Omega (\Omega^2-3 \Omega \omega_0 +2\omega_0^2)} \bigg]\left[\vec{n}(t)\times\hat{\vec{R}}_0\right] \,,\label{H2pGeneral}
\eeq
\end{widetext}
The first term describes the unperturbed dipole radiation, as we can see from a quick comparison with Eq.\eqref{EandMfields}. Once this term is expressed in the observer frame, $\vec{n}(t)$ also includes a term linear in the external perturbation, due to the first order Taylor expansion of the trigonometric functions in Eq.~\eqref{n_of_t}. The second term, that does not depend on $\vec{n}(t)$, is the only one that matters in the high frequency limit. The third term represents the modification to the dipole emission due to the external wave.

\subsubsection{Cross section}

\begin{figure}
\begin{center}
\includegraphics[scale=0.3]{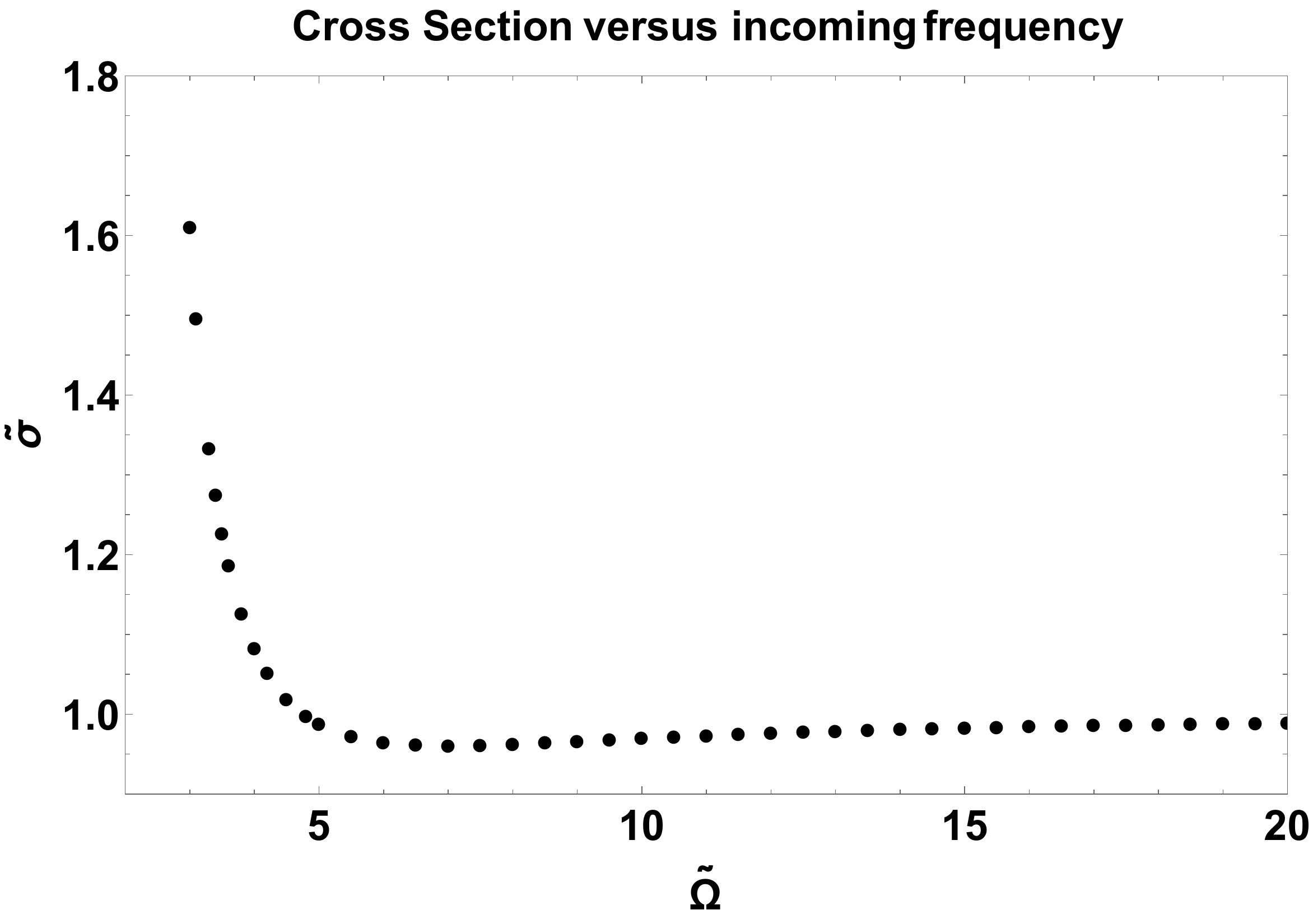}
\caption{Total scattering cross section for a dipolar wave as a function of the external frequency. Here, $\tilde{\sigma}\equiv \sigma/(A r_\circ^2)$ is shown as a function of $\tilde{\Omega}\equiv \Omega/\omega_0$. We set $c=1$ and the two angles $\gamma=\phi_0=0$. As expected, the cross section grows unboundedly for $\Omega=2\omega_0$. For large values of $\Omega$ we recover the standard high-frequency classical result~\eqref{EMOmegainfCrossSection}.}
\label{fig:cross_section}
\end{center}
\end{figure}

The scattering cross section is defined as the ratio between the energy emitted by the system in any given direction per unit of time, to the energy flux density of the incident radiation per unit of time. Considering that $\mathrm{d}I$ is the energy radiated per second by the binary into the solid angle $\mathrm{d}o$, we can define the differential cross section as
\begin{equation}
\mathrm{d}\sigma=\frac{\mathrm{d}I_{\rm scat}}{S_{\Omega}},
\end{equation}
where $S_\Omega$ is the modulus of the Poynting vector of the incoming wave.
Using the relation between intensity and Poynting vector and considering that the Poynting vector module is a time-varying quantity, we get
\begin{equation}
\frac{\mathrm{d}\sigma}{\mathrm{d}o} =\frac{\langle S_{\rm scat}\rangle R_0^2}{\langle S_{\Omega}\rangle},
\end{equation}
where the triangle brackets indicate a time average over one (or more) period and $\mathrm{d}o$ is the solid angle element given, with our choice of $\hat{\vec{R}}_0$, by
\begin{equation}
\mathrm{d}o=\cos\xi \mathrm{d}\xi \mathrm{d}\delta, \ \text{ with}\ \ \xi=[-\pi/2,\pi/2];\,\delta=[0,2\pi].
\end{equation}
In the high frequency limit, since the incoming wave is a monochromatic plane wave, its Poynting vector is
\begin{equation}
\vec{S}_{\Omega}=\left(\frac{c}{4\pi}E_{\Omega}^2 c^2_{\Omega t} \right){\mathbf{N}}\,.
\end{equation}
Its absolute value, averaged over one period of the EM wave ($2\pi/\Omega=T_{\Omega}$), is
\begin{equation}\label{Poynting_2p}
\langle S_{\Omega}\rangle=\frac{\Omega}{2\pi} \int_{T_{\Omega}}{S}_{\Omega} \mathrm{d}t= \frac{c E_{\Omega}^2}{8\pi}\,.
\end{equation}
To evaluate the Poynting vector of the scattered radiation we need to use the fields obtained in~\eqref{E2pGeneral} and~\eqref{H2pGeneral},
\begin{equation}
\langle S_{\rm scat}\rangle=\langle\frac{c}{4\pi}(\vec{E}_{2p}\times \vec{H}_{2p})\rangle=\frac{\Omega}{2\pi} \int_{T_{\Omega}}\frac{c}{4\pi}\lvert\vec{E}_{2p}\times \vec{H}_{2p}\rvert \mathrm{d}t\,.
\end{equation}
In the high frequency limit we can evaluate the differential scattering cross section using only the second term in Eq.~\eqref{E2pGeneral} and Eq.~\eqref{H2pGeneral},
\begin{equation}
\frac{\mathrm{d}\sigma}{\mathrm{d}o}=\left(\frac{q^2}{c^2 M}\right)^2 \left(2  c_{2 \xi} c^2_{\gamma -\delta}+c_{2 (\gamma -\delta )}-3\right)\,.
\end{equation}
The total high frequency scattering cross section is found by integrating the above, and yields
\begin{equation}\label{eq:thomson}
\sigma=\frac{32 \pi}{3}\left(\frac{q^2}{c^2 M}\right)^2\,,
\end{equation}
the standard Thomson result~\cite{Landau:1982dva}. Notice that $q^2/Mc^2$ is the classical charge radius. In the case of a circular orbit, the cross section in~\eqref{eq:thomson} can be given as a function of the unperturbed orbital frequency through~\eqref{omega_zero},
\beq
\sigma=\frac{32 \pi}{3}\left(\frac{r_{\circ}^3 \omega_0^2}{2 c^2}\right)^2= A r_\circ^2\,\label{EMOmegainfCrossSection},
\eeq
where $A=\frac{8 \pi}{3}\left(\frac{r_{\circ}^2 \omega_0^2}{c^2}\right)^2$. The total cross section for a wave with a generic frequency $\Omega$ is shown in Fig.~\ref{fig:cross_section} at $\Omega\gtrsim \omega_0$. 
From this plot, we can see that the cross section goes to infinity when the incoming frequency approaches twice of the orbital frequency and that in the high frequency limit it reaches the value given in Eq.~\eqref{EMOmegainfCrossSection}.

\section{Scattering of scalar waves}\label{Scattering of scalar waves}

For completeness, we now show that the previous results are straightforward to extend in the presence of a scalar interaction. Let's suppose that a binary system, made of two point-like scalar charges, interacts with a scalar wave and that is on a circular orbit of frequency $\omega_0$.

We consider the following action,
\begin{equation}
S=\int \mathrm{d}^4x \left[-\frac{1}{2}\partial^{\mu}\phi\partial_{\mu}\phi-\rho_q \phi\right]+S_m\,,
\end{equation}
where $\phi$ is the scalar field and $\rho_q$ is a general scalar charge density. The action for a free particle $S_m$ is as in \eqref{action1}. The scalar field is then governed by the Klein-Gordon equation
\begin{equation}\label{scalarfieldeq}
\partial^{\mu}\partial_{\mu} \phi=\rho_q\,.
\end{equation}
For a point-like charge $\rho_q=q\delta^3(\vec{x}-\vec{x}')  \frac{\mathrm{d}\tau}{\mathrm{d}t}$, in the non-relativist limit the static solution is given by the Green's function of the Laplace operator,
\begin{equation}
\phi(\vec{x})=-\frac{q}{4\pi}\frac{1}{\lvert\vec{x}-\vec{x}'\rvert}\,,\label{scalarpotential}
\end{equation} 
therefore, a Coulomb-like potential. From the Euler-Lagrange equation in a non relativistic regime, one finds,
\begin{equation}
m_1 a_1^j=-q_1\left(\frac{\partial \phi_2}{\partial x_j}\right)_1\,,
m_2 a_2^j=-q_2\left(\frac{\partial \phi_1}{\partial x_j}\right)_2\,,
\end{equation}
in which the scalar field $\phi_i$ is the potential produced by the particle $i$.

Let's now consider the case in which a scalar wave impinges on the system. We call $\delta\phi=\phi_{\Omega}$ the perturbation in the scalar potential. The equations of motion are then altered to include the interaction of the binary with the wave,
\beq
m_1 a_1^j&=&-q_1\left(\frac{\partial \phi_2}{\partial x_j}\right)_1-q_1\left(\frac{\partial \phi_{\Omega}}{\partial x_j}\right)_1\,,\\
m_2 a_2^j&=&-q_2\left(\frac{\partial \phi_1}{\partial x_j^2}\right)_2-q_2\left(\frac{\partial \phi_{\Omega}}{\partial x_j}\right)_2\,.
\eeq
Using the explicit form of the potential \eqref{scalarpotential}, transforming the equations to the CM frame and assuming that the scalar field is homogeneous enough to be evaluated directly in the CM position, we find
\beq
\ddot{\vec{R}}_{CM}&=& -\left(\frac{q_1+q_2}{m}\right)\left(\frac{\partial \phi_{\Omega}}{\partial x_j}\right)_{CM}\,,\\
\ddot{\vec{r}}&=& -\frac{1}{\mu}\frac{q_1 q_2}{ \lvert r\lvert^2}\vec{n} -\left(\frac{q_1}{m_1}-\frac{q_2}{m_2}\right)\left(\frac{\partial \phi_{\Omega}}{\partial x_j}\right)_{CM}\,.
\eeq
These equations are formally equivalent to the EM counterpart, and no further calculation is necessary.
The existence of a background of light bosons is motivated by the problem of dark matter, see e.g. \cite{Battaglieri:2017aum}. Their influence in binary systems was recently described in \cite{Blas:2016ddr,Boskovic:2018rub,LopezNacir:2018epg}. The scattering that we described here may be of interest to refine these studies.

\section{Scattering of gravitational waves}\label{Scattering of gravitational wave}

We now want to evaluate the effect of an incoming gravitational wave on a binary system. The binary system is made of two compact stars modeled by two point particles of mass $m_1$ and $m_2$ orbiting at an orbital frequency $\omega_0$. We consider an incoming monochromatic GW propagating in the $z$ direction of a fixed basis $(\vec{e}_{x},\vec{e}_{y},\vec{e}_{z})$, at a frequency $\Omega$. In the transverse traceless (TT) gauge, $H_{ij}$, the waveform is
\beq\label{Hij}
H_{ij} = &&\mathcal{P}_{ij}^{kl}\left\{H_{+}\,\cos\left(\Omega\,t -k\,z\right)\,e_{kl}^{+}\right.\nonumber \\
&&~~~~~~~~~~\left.
 +H_{\times}\,\sin\left(\Omega\,t -k\,z\right)\,e_{kl}^{\times}\right\} \,,
\eeq
where $e_{ij}^{+}=\vec{e}_{x}\otimes\vec{e}_{x}-\vec{e}_{y}\otimes\vec{e}_{y}$ and $e_{ij}^{\times}=\vec{e}_{x}\otimes\vec{e}_{y}+\vec{e}_{y}\otimes\vec{e}_{x}$ are the two polarization states of the GW~\cite{Maggiore:1900zz}, $\mathcal{P}_{ijkl}=P_{ik}P_{jl}-\frac{1}{2}P_{ij}P_{kl}$ is the TT projection operator, with $P_{ij}=\delta_{ij}-N_{i}N_{j}$ the projection onto the plane orthogonal to $\mathbf{N}$. As the wave is a solution of $\square H_{ij}=0$, we have,
\be
k=\pm\frac{\Omega}{c}\,,
\ee
and $H_{+}=cst$ and  $H_{\times}=cst$. As a consequence, the spatial derivatives of $H_{ij}$ will be suppressed, $\partial_{k}H_{ij}=\mathcal{O}\left(\frac{1}{c}\right)$ (homogeneity condition), in particular $\partial_{j}H_{ij}=0$ (transverse), and we can assume that in the near zone of the compact binary one has $H_{ij}\sim (H_{ij})_{A}\sim (H_{jk})_{\mathrm{CM}}$. These conditions will be used in the following when we will perform a perturbative expansion of the solution.

\subsection{Post-Newtonian formalism}

\subsubsection{Einstein's equations}

We want to solve the Einstein equations
\begin{equation}
G^{\mu\nu} = \frac{8\pi G}{c^4} T^{\mu\nu}\,,
\end{equation}
where $G^{\mu\nu}$ is the Einstein tensor and $T^{\mu\nu}$ is the stress-energy tensor for point particles 
\begin{equation}\label{stress-energy tensor}
T^{\mu\nu}=\sum_{A=1,2}\frac{m_A}{\sqrt{-g}}\frac{v_A^{\mu}v_A^{\nu}}{\sqrt{-\left(g_{\rho\sigma}\right)_{A}\frac{v_{A}^{\rho}v_A^{\sigma}}{c^2}}}\delta^{(3)}(\mathbf{x}-\mathbf{x}_A)\,,
\end{equation}
where $g$ is the determinant of the metric $g_{\mu\nu}$. Defining the gothic metric $\mathfrak{g^{\mu\nu}}=\sqrt{-g}g^{\mu\nu}$ and the tensor $H^{\mu\alpha\nu\beta}=\mathfrak{g}^{\alpha\beta}\mathfrak{g}^{\mu\nu}-\mathfrak{g}^{\alpha\nu}\mathfrak{g}^{\beta\mu}$, we have the well-known identity~\cite{PoissonWill2014}
\begin{equation}\label{HtoG}
\partial_{\alpha\beta}H^{\mu\alpha\nu\beta} = (-g)\left(2G^{\mu\nu}+\frac{16 \pi G}{c^4} t^{\mu\nu}_{\mathrm{LL}}\right)\,,
\end{equation}
where $t^{\mu\nu}_{\mathrm{LL}}$ is the Landau-Lifshitz tensor~\cite{PoissonWill2014}. Next we define the gravitational field
\begin{equation}\label{def gothic}
l^{\mu\nu}=\mathfrak{g}^{\mu\nu}-\eta^{\mu\nu},
\end{equation}
where $\eta^{\mu\nu}=\mathrm{diag}(-1,1,1,1)$ is the Minkowski metric and we impose the harmonicity condition on the metric perturbation $l^{\mu\nu}$,
\begin{equation}
\partial_{\nu}l^{\mu\nu} = 0\,.
\end{equation}
Using~\eqref{HtoG}, we can rewrite the field equations as
\begin{equation}\label{relaxedEEqinH}
\square l^{\mu\nu}=\frac{16 \pi G}{c^4}T^{\mu\nu}+\Lambda^{\mu\nu}\,,
\end{equation}
where
\begin{equation}
\Lambda^{\mu\nu} = \frac{16 \pi G}{c^4} t^{\mu\nu}_{\mathrm{LL}} + \partial_{\rho}l^{\mu\sigma}\partial_{\sigma}l^{\nu\rho} -l_{\rho\sigma}\partial_{\rho\sigma}l^{\mu\nu}\,,
\end{equation}
is at least quadratic in the gravitational field. In our case, $l^{\mu\nu}$ is formed by two different terms, the perturbation $H^{\mu\nu}$ due to the incoming GW that we are superimposing on the original unperturbed gravitational field $h^{\mu\nu}$. Thus, at linear order we can write
\begin{equation}\label{h and htt}
l^{\mu\nu}=h^{\mu\nu}+H^{\mu\nu}\,.
\end{equation}
Since $\square H^{\mu\nu} =0$, the field equations~\eqref{relaxedEEqinH} can be rewritten as a d'Alembertian equation for $h^{\mu\nu}$,
\begin{equation}\label{eqnonso}
\begin{split}
\square h^{\mu\nu}=&\frac{16 \pi G}{c^4}T^{\mu\nu}[m,h^{\mu\nu},H^{\mu\nu}]+\Lambda^{\mu\nu}[h^{\mu\nu},h^{\mu\nu}]+\\
&+\Lambda^{\mu\nu}[h^{\mu\nu},H^{\mu\nu}]+\Lambda^{\mu\nu}[H^{\mu\nu},H^{\mu\nu}]\,.
\end{split}
\end{equation}
The last term in \eqref{eqnonso} can be neglected when considering only the dominant, linear order in $H^{\mu\nu}$ terms. 

\subsubsection{Post-Newtonian iteration}

We perform the post-Newtonian iteration of the field equations in harmonic coordinates in the near-zone of the isolated source. As we are only interested in the effect of the external perturbation on the binary dynamics, we only need the lowest order PN expansion. We parametrize the metric by the usual PN potentials, using the variable $h^{00ii}\equiv h^{00}+h^{ii}$,
\begin{align}\label{hmunuPot}
& h^{00ii}=-\frac{4V}{c^2} + \mathcal{O}(c^{-4})\,,\\
& h^{0i}=-\frac{4V^{i}}{c^3} + \mathcal{O}(c^{-5})\,,\\
& h^{ij}= \mathcal{O}(c^{-4})\,.
\end{align}
Each potential obeys a flat space-time d'Alembertian equation sourced by the lowest order potentials and by some matter energy density components. We get
\begin{align}\label{eqnsVVi}
& \square V = -4\pi G \Upsilon -H^{ab}\partial_{ab}h^{00ii} \,,\\
& \square V^{i} = -4\pi G \Upsilon^{i} -H^{ab}\partial_{ab}h^{00ii} +\frac{1}{c}\partial_{t}H^{ia}\partial_{a}h^{00ii} \,,
\end{align}
where we have defined
\begin{equation}
\Upsilon = \frac{T^{00}+T^{ii}}{c^{2}} \,, \qquad \text{ and } \quad \Upsilon^{i}=\frac{T^{0i}}{c} \,.
\end{equation}
The first terms in the r.h.s. of Eqs.~\eqref{eqnsVVi} are of compact support, while the other terms are of non-compact support. We solve these equations perturbatively and up to linear order in $H$. The zeroth order corresponds to the Newtonian term and the equation becomes $\Delta V = -4\pi G \Upsilon$, with $\Upsilon = m_{1}\delta_{1}+m_{2}\delta_{2}$, and thus
\begin{equation}
V_{N}=\frac{Gm_{1}}{r_{1}}+\frac{Gm_{2}}{r_{2}}\,.
\end{equation}
We now decompose $V$ in a Newtonian part and a contribution linear in $H$, $V=V_{N}+V_{h}$. Inserting it into Eq.~\eqref{eqnsVVi}, we find
\begin{equation}
V_{h} = \Delta^{-1}\left[-H^{ab}\partial_{ab}\left(\frac{Gm_{1}}{r_{1}}+\frac{Gm_{2}}{r_{2}}\right)\right]\,.
\end{equation}
Using the fact that $H^{ij}\sim(H^{ij})_{CM}$, we see that the inverse Laplacian will not act on $H$. Further commuting it with the spatial derivatives, we get
\begin{equation}\label{Vh}
V_{h} = \frac{Gm_{1}}{r_{1}}H_{ij}n_{1}^{i}n_{1}^{j}+\frac{Gm_{2}}{r_{2}}H_{ij}n_{2}^{i}n_{2}^{j}\,.
\end{equation}
The potential $V^{i}$ can be obtained in a similar way.

\subsubsection{Geodesic equation}

The geodesic equations for point-particles is equivalent to the conservation of the matter stress-energy tensor, $\nabla_{\nu}T^{\mu\nu}=0$. We express the resulting equations for particle 1 as~\cite{Blanchet:2013haa}
\begin{equation}
\frac{\mathrm{d}\left(P^{i}\right)_{1}}{\mathrm{d}t} = \left(F^{i}\right)_{1} \,,
\end{equation}
with
\begin{align}
& P^{i} = \frac{g^i_{\mu}v^{\mu}}{\sqrt{-g_{\rho\sigma}v^{\rho}v^{\sigma}}}\,, \\
& F^{i} = \frac{1}{2}\frac{\partial^{i}g_{\mu\nu}v^{\mu}v^{\nu}}{\sqrt{-g_{\rho\sigma}v^{\rho}v^{\sigma}}}\,.
\end{align}
Using the expression of the metric as a function of the potentials, see Eq.~\eqref{hmunuPot}, we obtain at linear order in $H$
\begin{align}
P_{1}^{i} &= v_{1}^{i} -v_{1}^{j}(H^i_{j})_{1},\\
F_{1}^{i} &= \frac{1}{2}v_{1}^{j}v_{1}^{k}(\partial_{i}H_{jk})_{1} +(\partial^{i}V)_{1} \,.
\end{align}
Using the relation $V=V_{N}+V_{h}$, with $V_{h}$ given by Eq.~\eqref{Vh}, we finally obtain the acceleration of particle $1$,
\begin{equation}\label{eqacceleration}
a^i_1=-\frac{G m_2 }{r_{12}^{2}}\left(1+\frac{3}{2}H_{jk}n_{12}^{j}n_{12}^{k}\right)n_{12}^{i} +\frac{\mathrm{d}H_{ij}}{\mathrm{d}t}v_1^{j}\,.
\end{equation}
Here we ignored all higher order post-Newtonian corrections, since they will be sub-leading in the computation of the cross-section.

\subsubsection{Lagrangian formulation}

The equations of motion~\eqref{eqacceleration} can be derived from the Lagrangian
\beq
L &=&\frac{G m_1 m_2}{r_{12}}\left(1+\frac{1}{2}H_{ij}n_{12}^{i}n_{12}^{j}\right) +\frac{1}{2}m_1 v_1^2\nonumber\\
&-&\frac{1}{2}m_1H_{ij}v_1^i v_1^j +\frac{1}{2}m_2 v_2^2-\frac{1}{2}m_2H_{ij}v_2^i v_2^j\,.\label{lagrangian}\nn
\eeq
Varying the Lagrangian with respect to the velocities, we obtain the linear momentum
\beq
P^i&\equiv& \sum_{A=1,2}\frac{\delta L}{\delta v_{A}^{i}}\nonumber\\
&=& m_1	v_1^i -m_1 H_{ij} v^j_1+m_2 v_2^i -m_2 H_{ij} v^j_2.\nonumber
\eeq
It is possible to see that the time-derivative of the momentum is zero. Then we get the energy associated with the binary motion,
\begin{equation}
\begin{split}
E \equiv& \sum_{A=1,2}\frac{\delta L}{\delta v_{A}^{i}}v_{A}^{i}-L\\
=&-\frac{G m_1 m_2}{r_{12}}\left(1+\frac{1}{2}H_{jk}n_{12}^{j}n_{12}^{k}\right)+\frac{1}{2}m_1 v_1^2\\
&-\frac{1}{2}m_1 H_{ij} v^i_1 v^j_1+\frac{1}{2}m_2 v_2^2-\frac{1}{2}m_2 H_{ij} v^i_2 v^j_2\,.
\end{split}
\end{equation}
Similarly the angular momentum $J^i$ is given by
\beq
&J^i&\equiv \epsilon^{i}_{jk}\sum_{A=1,2}x_{A}^{j}\frac{\delta L}{\delta v_{A}^{k}}\nonumber\\
& =&\epsilon_{ijk}\left[m_1\left(x_1^j v_1^k - H_{kl}x_1^j v_1^l\right)
+m_2\left(x_2^j v_2^k - H_{kl}x_2^j v_2^l\right)\right]\,,\nonumber
\eeq
where $\epsilon^{i}_{jk}$ is the Levi-Civita tensor. Finally, we define the center-of-mass integral
\be
G^{i} = m_{1}x_{1}^{i} - m_{1}H^i_{j}x_{1}^{j} + [1\leftrightarrow 2]\,.
\ee
The conservation laws associated with all these quantities are
\beq
\frac{\mathrm{d}P^{i}}{\mathrm{d}t}  &=& 0 \,,\\
\frac{\mathrm{d}G^{i}}{\mathrm{d}t}  &=& P^{i} - m_{1}\left(\dot{H}^i_{j}\right)_{1}x_{1}^{j}- m_{2}\left(\dot{H}^i_{j}\right)_{2}x_{2}^{j}\,,\\
\frac{\mathrm{d}E}{\mathrm{d}t} &=& \frac{1}{2} m_1\left(\dot{H}_{ij}\right)_{\mathrm{CM}}v_{1}^{i}v_{1}^{j} +\frac{1}{2} m_2\left(\dot{H}_{ij}\right)_{\mathrm{CM}}v_{2}^{i}v_{2}^{j}\nonumber\\
& -&\frac{Gm_{1}m_{2}}{2r_{12}}\left(\dot{H}_{ij}\right)_{\mathrm{CM}}n_{12}^{i}n_{12}^{j}\,,\\
\frac{\mathrm{d}J^{i}}{\mathrm{d}t} &=& \epsilon_{ijk}\bigg[-m_1v_{1}^{j}\left(H_{km}\right)_{\mathrm{CM}}v_{1}^{m} -m_2v_{2}^{j}\left(H_{km}\right)_{\mathrm{CM}}v_{2}^{m}\nonumber\\
&+&\frac{Gm_{1}m_{2}}{r_{12}}n_{12}^{j}\left(H_{km}\right)_{\mathrm{CM}}n_{12}^{m} \bigg]\,.
\eeq
Unlike the Newtonian result, these quantities are not conserved, due to the incoming GW; the only conserved quantity here is the momentum $P^i$. 

We now wish to work in the center-of-mass coordinates. We define the total mass $m$, the symmetric mass ratio $\nu$ and the relative position $x^i$ and velocity $v^i$ as
\beq
m&=&m_1+m_2\,,\\
\nu&=&\frac{\mu}{m}=\frac{m_{1}m_{2}}{m^{2}}\,,\\
x^{i}&=&x_{1}^{i}-x_{2}^{i},\,\quad r=\vert\mathbf{x}\vert\,,\\
v^{i}&=&v_{1}^{i}-v_{2}^{i}\,,\ a^{i}\equiv a_{1}^{i}-a_{2}^{i}\,.  
\eeq
The center of mass coordinates are obtained by solving the equation
\be
G^{i}=0 \,.
\ee
It implies the well-known Newtonian results, that are still valid at linear order in $H_{ij}$,
\begin{align}
& x_{1\,,\mathrm{CM}}^{i}=\frac{m_{2}}{m}x^{i}\,,\qquad   x_{2\,,\mathrm{CM}}^{i}=-\frac{m_{1}}{m}x^{i},\\
& v_{1\,,\mathrm{CM}}^{i}=\frac{m_{2}}{m}v^{i}\,,\qquad   v_{2\,,\mathrm{CM}}^{i}=-\frac{m_{1}}{m}v^{i}\,.
\end{align}
In the center of mass coordinates, the relative acceleration is given by
\begin{align}
a^{i} = -\frac{Gm}{r^{2}}\left( 1+\frac{3}{2}H_{jk}n^{j}n^{k} \right)n^{i} +\dot{H}_{ij}v^{j}\,,
\end{align}
and the conservation laws are now
\beq
\frac{\mathrm{d}P^{i}}{\mathrm{d}t}  &=& 0 \,,\\
\frac{\mathrm{d}E}{\mathrm{d}t} &=&  \frac{1}{2} m\nu\dot{H}_{ij}v^{i}v^{j} -\frac{Gm^{2}\nu}{2r}\dot{H}_{ij}n^{i}n^{j} \,,\\
\frac{\mathrm{d}J^{i}}{\mathrm{d}t}& = & \epsilon_{ijk}m\nu\bigg[ \frac{Gm}{r}n^{j}H_{km}n^{m}
-v^{j}H_{km}v^{m} \bigg] \,.
\label{ConsLaw}
\eeq
%
\subsection{Hamiltonian formulation and angle-action variables}

To understand the gravitational problem, we will follow a different route from that we used in the electromagnetic example.
We will use an approach based on angle-action variables. The dynamics of a Keplerian orbit in Delaunay variables is well known, and we will use the powerful tool of perturbation theory in angle-action variables to describe the evolution of the perturbed system~\cite{BinneyTremaine}. The advantages of such an approach is that the calculations are simpler, notably because they capture the symmetries of the system. By promoting the integrals of motion to coordinate variables in the phase space, the dynamics of the system becomes very simple, as we will see. In particular, it allows a simple treatment of generic orbits and of the resonances that occur in such systems. However, in this work we will focus mostly on circular orbits and resonances will be absent from the final result. In Appendix~\ref{angleaction}, we review the Hamiltonian in the Delaunay variables, and explain the basics of perturbation theory in angle-action variables.

\subsubsection{Hamiltonian in the modified Delaunay variables}

The first step is to determine the Hamiltonian from the perturbed Lagrangian, and then to express it as a function of the modified Delaunay variables $\left(\theta_{1,2,3},\,J_{1,2,3}\right)$ (see Appendix~\ref{angleaction1}). We start from the reduced perturbed Lagrangian in the center-of-mass coordinates, in spherical coordinates,
\beq
\tilde{L} &= & \frac{G m}{r}\left[1+\frac{1}{2}\,H_{ij}n^{i}n^{j}\right] + \frac{1}{2}\,\dot{r}^{2}\left[1-H_{ij}n^{i}n^{j}\right]\nonumber\\
&+& \frac{1}{2}\,r^{2}\,\dot{\theta}^{2}\left[1-H_{ij}\theta^{i}\theta^{j}\right] + \frac{1}{2}\,r^{2}\,\sin^{2}\theta\,\dot{\varphi}^{2}\left[1-H_{ij}\varphi^{i}\varphi^{j}\right]\nonumber \\
& -& r\,\dot{r}\,\dot{\theta}\,H_{ij}n^{i}\theta^{j} - r\,\sin\theta\,\dot{r}\,\dot{\varphi}\,H_{ij}n^{i}\varphi^{j}\nonumber\\
& -& r^{2}\,\sin\theta\,\dot{\theta}\,\dot{\varphi}\,H_{ij}\theta^{i}\varphi^{j} \,,
\eeq
we derive the conjugate momenta $p_{x}=\partial\tilde{L}/\partial\dot{x}$,
\beq
p_{r}&=&\dot{r}\left[1-H_{ij}n^{i}n^{j}\right] - r\,\dot{\theta}\,H_{ij}n^{i}\theta^{j} - r\,\sin\theta\,\dot{\varphi}\,H_{ij}n^{i}\varphi^{j} \,,\nonumber\\
p_{\theta}&=&r^{2}\dot{\theta}\left[1-H_{ij}\theta^{i}\theta^{j}\right] - r\,\dot{r}\,H_{ij}n^{i}\theta^{j} - r^{2}\,\sin\theta\,\dot{\varphi}\,H_{ij}\theta^{i}\varphi^{j} \,,\nonumber\\
p_{\varphi}&=&r^{2}\sin^2\theta\,\dot{\varphi}\left[1-H_{ij}\varphi^{i}\varphi^{j}\right] - r\,\sin\theta\,\dot{r}\,H_{ij}n^{i}\varphi^{j} \nonumber\\
&-& r^{2}\,\sin\theta\,\dot{\theta}\,H_{ij}\theta^{i}\varphi^{j} \,,
\eeq
and then the reduced perturbed Hamiltonian
\beq
\mathcal{\tilde{H}} & \equiv& p_{r}\dot{r}+p_{\theta}\dot{\theta}+p_{\varphi}\dot{\varphi}-\tilde{L} \nonumber\\
& =& -\frac{G m}{r}\left[1+\frac{1}{2}\,H_{ij}n^{i}n^{j}\right] + \frac{1}{2}p_{r}^{2}\left[1+H_{ij}n^{i}n^{j}\right]\nonumber\\
& +& \frac{1}{2r^{2}}p_{\theta}^{2}\left[1+H_{ij}\theta^{i}\theta^{j}\right] + \frac{1}{2r^{2}\sin^2\theta}p_{\varphi}^{2}\left[1+H_{ij}\varphi^{i}\varphi^{j}\right] \nonumber\\
&+& \frac{p_{r}\,p_{\theta}}{r}\,H_{ij}n^{i}\theta^{j} + \frac{p_{r}\,p_{\varphi}}{r\sin\theta}\,H_{ij}n^{i}\varphi^{j}  \nonumber\\
&+&\frac{p_{\theta}\,p_{\varphi}}{\rho^{2}\sin\theta}\,H_{ij}\theta^{i}\varphi^{j}\,.
\eeq
The perturbed Hamiltonian depends explicitly on time through the perturbation $H_{ij}$, given by Eq.~\eqref{Hij}.

The next step is to write the Hamiltonian as a function of the modified Delaunay variables. This can only be achieved with an expansion in the eccentricity $e$. In the following we will only consider the perturbation of a circular orbit. At this order we have that $J_{2}=0$ and $\theta_{2}$ is not defined. Our new set of variables is thus $\left\{\theta_{1,3},J_{1,3}\right\}$, and the relations between the old canonical variables and the angle-action ones are
\beq
r&=&\frac{J_{3}^{2}}{G m}\,,\qquad \theta=\frac{\pi}{2}-\arccos\left[1-\frac{J_{1}}{J_{3}}\right]\,, \nonumber\\
\varphi&=&-\theta_{1}+\theta_{3}\,,\nonumber\\
p_{r}&=&0\,,\qquad p_{\theta}=0\,,\qquad p_{\varphi}=J_{3}-J_{1}\,.
\eeq
The Hamiltonian that arises out of this procedure is
\be
\tilde{\tilde{\mathcal{H}}}=-\frac{G^{2}m^{2}}{2J_{3}^{2}}\left[1+\left(H_{ij}n^{i}n^{j}\right)-\left(H_{ij}\lambda^{i}\lambda^{j}\right)\right] +\Omega\mathcal{T} \,,
\ee
where we have introduced a new variable $\tau$ and its conjugate $\mathcal{T}$ to absorb the explicit dependence in time, cf. App.~\ref{angleaction}. In particular it depends not only on the actions but also on the angle variables. The dependence on the variable $\tau$ is only through the incoming GW and the only modes that contribute to the Fourier expansion are $k_{\tau}=\pm 1$. Then, as $\boldsymbol{\Omega}_{0}\left(\mathbf{J}\right) = \frac{\mathrm{d}\tilde{\tilde{\mathcal{H}}}_{0}}{\mathrm{d}\mathbf{J}} = \left(0,0,\frac{G^{2}m^{2}}{J_{3}^{3}},\Omega\right)$, we can see that the resonance occurs when
\be
\Omega = \pm k_{3} n\,,
\ee
where $n=\frac{G^{2}m^{2}}{J_{3}^{3}}$ is the orbital frequency of the binary and $k_{3}\in\mathbb{N}$.

We can also use the new set of angle-action variables $\left(\boldsymbol{\theta}',\,\boldsymbol{J}'\right)$, as constructed in App.~\ref{angleaction2}. The Hamiltonian for this set of variables is
\be
\tilde{\mathcal{H}}'\left(\mathbf{J}'\right) = -\frac{G^{2}m^{2}}{{J'_{3}}^{2}} +\Omega\,\mathcal{T}' \,.
\ee
The Hamilton equations are then
\begin{align}
& \dot{\mathbf{J}'} = 0 \,,\\
& \dot{\theta'_{1}} = 0\,, \quad \dot{\theta'_{3}}=\frac{G^{2}m^{2}}{{J'_{3}}^{3}} \quad \text{ and }\quad \dot{\tau'} = \Omega \,.
\end{align}

\subsubsection{Variation of the orbit elements}

We now relate the new set of variables to the orbit elements and obtain their evolution. From $J_{3}=\sqrt{Gma}$, we get
\be
\frac{\mathrm{d}a}{\mathrm{d}t} = 2\sqrt{\frac{a}{Gm}}\frac{\mathrm{d}J_{3}}{\mathrm{d}t} = -2\sqrt{\frac{a}{Gm}}\frac{\partial\tilde{\mathcal{H}}_{1}}{\partial\theta_{3}}\,.
\ee
From $J_{1}=J_{3}\left(1-\cos\iota\right)$ we get, using the previous relation,
\be
\frac{\mathrm{d}\iota}{\mathrm{d}t} = \frac{\cos\iota-1}{\sin\iota}\frac{1}{a}\frac{\mathrm{d}a}{\mathrm{d}t}\,.
\ee
We also have
\be
\frac{\mathrm{d}\psi}{\mathrm{d}t} = -\frac{\mathrm{d}\theta_{1}}{\mathrm{d}t}\,, \quad \frac{\mathrm{d}\omega}{\mathrm{d}t} =-\frac{\mathrm{d}\psi}{\mathrm{d}t} \,, \text{ and }\quad \frac{\mathrm{d}l}{\mathrm{d}t} = \frac{\mathrm{d}\theta_{3}}{\mathrm{d}t}\,.\nonumber
\ee
To obtain explicit results we specify to some specific configurations.

\paragraph{Parallel to the orbital plane: $\alpha=0$, $\beta=\pi/2$, $\kappa=\pi/2+\iota$}

We compute the variation of the energy, defined as $E\equiv H$, and get
\beq\label{dEdt1}\nn
\frac{J_{3}^2}{G^2 m^2 \Omega}\frac{\mathrm{d}E}{\mathrm{d}t} &=& -\frac{H_{\times}}{2} \sin(2\iota) \cos(\Omega t) \sin(2\zeta) \\
&& \hspace{-.2cm}+\frac{H_{+}}{32} \sin(\Omega t) \left(\cos(4\iota)-17\right) \cos(2\zeta).\  \quad
\eeq
The variation of the semi-major axis is given by
\beq\nn
\sqrt{\frac{a}{Gm}}\frac{\mathrm{d}a}{\mathrm{d}t} &=& -\frac{H_{+}}{8} \cos(\Omega t) \left(\cos(4\iota)-17\right)\sin(2\zeta) \\
&& + 2 H_{\times} \sin(2\iota) \sin(\Omega t)\cos(2\zeta) ,\ 
\eeq
while the variation of the inclination angle is
\beq\nn
\frac{a^{3/2}}{\sqrt{G m}}\frac{\mathrm{d}\iota}{\mathrm{d}t} &=& 2H_{\times} (1-\cos\iota)\cos\iota \sin(\Omega t) \cos(2\zeta) \\
&& \hspace{-.7cm}-\frac{H_{+}}{16} \cos(\Omega t) (\cos (4\iota)-17) \tan \left(\frac{\iota}{2}\right) \sin(2\zeta).\  \qquad
\eeq
%

\paragraph{Perpendicular to the orbital plane: $\alpha=0$, $\beta=\pi/2$, $\kappa=\iota$}
The variation of the energy is
\beq\label{dEdt2}\nn
\frac{J_{3}^2}{G^2 m^2 \Omega}\frac{\mathrm{d}E}{\mathrm{d}t} &=& -\frac{H_{+}}{16} \sin (\Omega t) (\cos (2\iota)+3)^2 \cos(2\zeta) \\
&& - H_{\times} \cos(\Omega t) \cos^2\iota \sin(2\zeta) .
\eeq
The variation of the semi-major axis is given by
\beq\nn
-\sqrt{\frac{a}{Gm}}\frac{\mathrm{d}a}{\mathrm{d}t} &=& 4H_{\times} \sin(\Omega t) \cos^2\iota\cos(2\zeta) \\
&&  \hspace{-.7cm}+\frac{H_{+}}{4} \cos(\Omega t) (\cos (2\iota)+3)^2 \sin(2\zeta) ,
\eeq
and the variation of the inclination angle is
\beq\nn
\frac{a^{3/2}}{\sqrt{G m}}\frac{\mathrm{d}\iota}{\mathrm{d}t}& =& 2H_{\times} \cos^{2}\iota \tan\left(\frac{\iota}{2}\right) \sin(\Omega t) \cos(2\zeta) \\
&&  \hspace{-.7cm}+\frac{H_{+}}{8}\cos (\Omega t) (\cos (2\iota)+3)^2 \tan \left(\frac{\iota}{2}\right) \sin(2\zeta) .\  \qquad
\eeq

\subsubsection{Scattered gravitational wave}
The asymptotic waveform is given by~\cite{Blanchet:2013haa},
\begin{widetext}
\be
l_{km}^{TT} = \frac{2G}{c^2R}\mathcal{P}_{ijkm}\sum_{l=2}^{\infty}\frac{1}{c^l l!}\left\{N_{L-2}U_{ijL-2}\left(T-R/C\right)-\frac{2l}{(l+1)c}N_{aL-2}\varepsilon_{ab(i}V_{j)bL-2}\left(T-R/C\right)\right\} + \mathcal{O}\left(\frac{1}{R^2}\right) \,,
\ee
\end{widetext}
where $\left(T,\,R\right)$ are the radiative coordinates. We recall that $\mathcal{P}_{ijkm}$ is the TT projection and the radiative moments $U_{L}$ and $V_{L}$ are related to the mass-type and current-type moments of the source. Here we suppose that the same relation still holds at linear order in $H_{ij}$, that is, $U_{L}(T)=M_{L}^{(l)}(T)$. The waveform is then given at our order by
\be\label{GWtotal}
l_{km}^{TT} = H^{\mathrm{TT}}_{km} + \frac{2G}{c^4R}\mathcal{P}_{ijkm}M_{ij}^{(2)} + \mathcal{O}\left(\frac{1}{R^2}\right)\,,
\ee
The contribution from the incoming GW, $H^{\mathrm{TT}}_{ij}$, has to be expanded at future null infinity, i.e. when $R\rightarrow+\infty$ keeping $T-\frac{R}{c}$ constant. We get\footnote{We did not prove this expression and only consider the general structure of a gravitational wave at infinity}
\beq
\frac{2\Omega R}{c}H^{\mathrm{TT}}_{ij}&=&\mathcal{P}_{ijkl}\biggl[H_{+}e_{ij}^{+}\sin\left[\Omega \left(T-\frac{R}{c}\right)\right] \nonumber\\
&&~~~~~~-H_{\times}e_{ij}^{\times}\cos\left[\Omega \left(T-\frac{R}{c}\right)\right]\biggr]\,.\nonumber
\eeq
Then, we have to link the canonical moments $M_{L}$ and $S_{L}$ to the real source moments $I_{L}$ and $J_{L}$, and then to figure out the expression of these source moments. We have the relation
\be
M_{L}(t)=I_{L}(t)+\delta I_{L}(t)\,,
\ee
In our case we are only interested by $M_{ij}$,  for which $\delta I_{ij}=0$. The source dipole moment is given by the usual formula,
\beq
I_{ij}(t) &=&\mathcal{FP}_{B=0}\int\mathrm{d}^{3}\mathbf{y}|\mathbf{y}|^{B}\Big[ \nonumber\\
&&~~~~~\left.\int_{-1}^{1}\mathrm{d}z \delta_{2}(z)\hat{y}_{ij}\bar{\Sigma}\left(\mathbf{y},t \right)\right] + \mathcal{O}({c^{-2}})\,,
\eeq
where $\mathcal FP$ is the finite part \cite{Blanchet:2013haa}. 
After some calculation we obtain,
\ba
M_{ij} &=& I_{ij} +\mathcal{O}\left(\frac{1}{c^2}\right)\,,\\
I_{ij} &=& m_{1}y_{1}^{\langle ij\rangle} +m_{2}y_{2}^{\langle ij\rangle} -\frac{m_{1}}{7}\left(H^{\langle ij\rangle}y_{1}^{2} +4H^{\langle i}_{a}y_{1}^{j\rangle}y_{1}^{a}\right) \nonumber\\
&-&\frac{m_{2}}{7}\left(H^{\langle ij\rangle}y_{2}^{2} +4H^{\langle i}_{a}y_{2}^{j\rangle}y_{2}^{a}\right) \,.
\ea
The second term in the gravitational waveform~\eqref{GWtotal} is given by
\be
h_{km}^{\mathrm{TT}} = \frac{2G}{c^4 R} \mathcal{P}_{ijkm} I_{ij}^{(2)} \,.
\ee
The projection onto the plus and cross polarizations gives,
\ba
h_{+} &=& \frac{G}{c^4 R}\left(P_{i}P_{j}-Q_{i}Q_{j}\right)I_{ij}^{(2)} \,, \\
h_{\times} &=& \frac{G}{c^4 R}\left(P_{i}Q_{j}+Q_{i}P_{j}\right)I_{ij}^{(2)}\,.
\ea
The explicit expression of the polarizations is given in the Appendix~\ref{app:polarization}. We can see that the amplitude of the scattered wave scales as $\omega_0^{-4/3}$.

\subsection*{The energy balance equation}
Before going further, we want to check that the energy balance equation is verified,
\be
\langle\frac{\mathrm{d}E}{\mathrm{d}t}\rangle = -\langle\mathcal{F}\rangle\,,
\ee
where the brackets stand for the angular average over one orbital period and the left-hand side has been computed in Eqs.~\eqref{dEdt1}-\eqref{dEdt2}. The gravitational flux is defined as,
\beq\nn
\mathcal{F} &\equiv & c\int_{\mathcal{S}}(-g)t_{\mathrm{LL}}^{0i}\mathrm{d}S_{i} \\
&=& \frac{c^{3}}{16\pi G}\int_{\mathcal{S}} \partial_{\tau}H_{ij}\partial_{\tau}h^{ij}\,R^{2}\mathrm{d}\Sigma\,,
\eeq
where $\mathrm{d}S_{i}=N^{i}R^{2}\mathrm{d}\Sigma$ is the surface element of the two-dimensional surface $\mathcal{S}$. Inserting the expressions for $H^{\mathrm{TT}}_{ij}$ and $h^{\mathrm{TT}}_{ij}$ we get for the first configuration,
\beq
\mathcal{F} &=& \frac{128 a^2 \nu m \omega_{0}^6 H_{+} \sin ^3\left(\frac{\pi\Omega }{\omega_{0}}\right) \cos \left(\frac{\pi  \Omega }{\omega_{0}}\right)}{15\pi \Omega \left(\Omega^2-4\omega_{0}^2\right)} ,
\eeq
and for the second configuration,
\beq
\mathcal{F} &=& \frac{8 a^2 \nu  m \omega_{0}^5 \sin ^3\left(\frac{\pi  \Omega }{\omega_{0}}\right) \cos \left(\frac{\pi  \Omega }{\omega_{0}}\right) (14 H_{+} \omega_{0}-5 H_{\times} \Omega )}{15 \pi \Omega \left(\Omega ^2-4 \omega_{0}^2 \right)} .\ \qquad
\eeq
After averaging over one orbital period the variation of the energy~\eqref{dEdt1}-\eqref{dEdt2}, we can see that the quadrupole formula is respected,
\be
\langle\frac{\mathrm{d}E'}{\mathrm{d}t}\rangle = -\langle\mathcal{F}\rangle\,,
\ee
where we have defined the modified energy
\beq\nn
E' &=& E +\frac{Gm^2\nu}{2r}\left(H_{ij}n^{i}n^{j}\right)+\frac{1}{2}m\nu\left(H_{ij}v^{i}v^{j}\right)\\
&=& -\frac{Gm^2\nu}{2r}\left[1-\frac{1}{2}\left(H_{ij}n^{i}n^{j}\right)\right]+\frac{1}{2}m v^{2}\,.
\eeq

\subsection*{The cross section}

Using the previous results, the scattering cross section can be computed generically for any binary orientation and any incoming gravitational wave. We find that, for a $+$-polarized wave, for instance, the scattering cross section depends on the angle with which the wave hits the binary. However, in the limit when $\Omega\rightarrow\infty$, the cross section for either edge- or head-on configurations is the same,
\be
\sigma = \frac{11408\,\pi\nu^2}{2205}\left(\frac{v}{c}\right)^4\frac{a^6}{\lambda_{GW}^4} \,.\label{crosssection1}
\ee
where we have introduced the orbital radius of the binary system, $a=\frac{v}{\omega_0}$, as well as the wavelength of the incoming GW, $\lambda_{GW}\equiv\frac{c}{\Omega}$.
From this formula we can see that at equal total mass, the effect is larger the slower the system. Also, it highlights a structure similar to that of Rayleigh scattering of light. 

Note that the cross section is not allowed to grow unboundedly, since that would take us away from the perturbative regime we work in. In particular, we must require that the scattered wave is always of much smaller amplitude than the incoming GW. Evaluating the scattered wave a wavelength away from the scatterer, we find that the cross-section can be expressed as
\be\label{crosssection2}
\sigma=C_1^2\,\frac{11408\,\pi\,a^2}{2205}\,,
\ee
with $C_1\ll 1$ the ratio between scattered and incident GWs. In addition, we must require that $C_2\equiv H\Omega/\omega_0\ll 1$, to ensure that the back-reaction on the binary is small. This condition prevents the cross section from getting arbitrarily large for weakly-bound binaries.

\section{Conclusions}

After a period of consolidation of the detection of 
gravitational waves, it is likely that we will enter a period of {\it precision gravitational wave physics}. This will require a better control on the possible effects that affect the production and propagation of gravitational waves. This will also be essential to use gravitational waves as an ultimate prove of the constituents of the Universe. In this work, we have extended previous calculations of propagation of gravitational waves by studying an important scattering system: a binary of compact objects.

Our results show that gravitationally-bound binaries are able to scatter incoming gravitational radiation. We have computed the scattered field for general configurations, two special cases are shown in Appendix~\ref{app:polarization}. 
Consider now the binary neutron-star systems: PSR J1411+2551~\cite{Martinez:2017jbp} (orbital period $2.6$ days, total mass $2.5\mathrm{M}_{\odot}$) and the ultra-relativistic pulsar PSR J1946+2052~\cite{Stovall:2018ouw} (orbital period $0.078$ days, total mass $2.5\mathrm{M}_{\odot}$). For a GW incoming at a frequency $f=\Omega/(2\pi)=200\mathrm{Hz}$, we find from the expressions in App.~\ref{app:polarization} that the order of magnitude of the correction in the amplitude -- due to scattering -- is, for the first binary pulsar $h_{+,\times}\sim 10^{-5}\,\mathrm{H}_{+,\times}$, while for the ultra-relativistic pulsar we have $h_{+,\times}\sim 10^{-7}\,\mathrm{H}_{+,\times}$. As expected from Eq.~\eqref{crosssection2} the effect is larger for slowly rotating pulsars. We also find that the effect is slightly stronger for the configuration of an incoming gravitational wave parallel to the orbital plane.

The numbers above are small, but not desperately small as to be discouraging. However, how likely is such an event? The magnitude of the scattered wave is insensitive (to this order) to the structure of the compact objects forming the binary. Thus, stellar mass black-hole binaries of similar periods will give rise to similar scattered amplitudes. The number of stellar mass black holes in the central parsec of our own Galactic nucleus was recently reported to be significant, of order $N={\cal O} (10^4)$~\cite{2018Natur.556...70H}. If this finding generalizes to other supermassive black holes in galactic nuclei, this implies that there exists a substantial screen of potential scatterers around supermassive black holes. Such screen may give rise to detectable levels of scattered radiation, either from binaries (the object of the current study) or from isolated objects (e.g. Ref.~\cite{McKernan:2014hha}). For example, consider GWs generated by stellar-mass black hole binaries in the last stages of the inspiral and merger, and (barely) detectable by LIGO, $f=\Omega/(2\pi) \sim$ 20 {\rm Hz}. The emitting-binary is close to the galactic center, and the emitted GWs will now cross a screen of binaries, which we assume have parameters close to the binary pulsar above (orbital period $2.6$ days, total mass $2.5\mathrm{M}_{\odot}$).
Using the scattering cross-section~\eqref{crosssection1}, and the number density found in the galactic center $n$, the mean free path of the GW is $\sim 1/(n\sigma)\sim 100\,{\rm Mpc}$, a number which is clearly too big to be of relevance. These estimates assume quasi-circular motion, we do not expect any qualitatively important change to occur when eccentricity is included.

\begin{acknowledgments}
We are indebted to Perimeter Institute for hosting us while this work started,
and to the Yukawa Institute for hospitality in the last stages of the work.
The authors acknowledge financial support provided under the European Union's H2020 ERC Consolidator Grant ``Matter and strong-field gravity: New frontiers in Einstein's theory'' grant agreement no. MaGRaTh--646597.
This project has received funding from the European Union's Horizon 2020 research and innovation programme under the Marie Sklodowska-Curie grant agreement No 690904.
The authors would like to acknowledge networking support by the GWverse COST Action CA16104 ``Black holes, gravitational waves and fundamental physics''. L.A. acknowledges financial support provided by Funda\c{c}ao para a Ci\^{e}ncia e a Tecnologia Grant number PD/BD/128232/2016 awarded in the framework of the Doctoral Programme IDPASC-Portugal.
\end{acknowledgments}

\clearpage
\newpage
\appendix
\section{Radiation from the CM}\label{Radiation from the CM}
The scalar and vector potentials produced by one accelerated particle are given by the Li\'enard–Wiechert potentials,
\begin{equation}
\Phi=\frac{q}{\left( {R}_0 -\frac{(\vec{v}\cdot \hat{\vec{R}}_0){R}_0}{c}\right)},\,\,\, \vec{A}=\frac{q \vec{v}}{c\left( {R}_0 -\frac{(\vec{v}\cdot \hat{\vec{R}}_0){R}_0}{c}\right)}\,.
\end{equation}
We can then find the electric and magnetic field for the accelerated charge in a relativistic context. For small velocities we get,
\beq
\vec{E}&=&\frac{q}{{R_0}^2}\hat{\vec{R}}_0+\frac{q}{c^2  {R_0}}\hat{\vec{R}}_0\times(\hat{\vec{R}}_0 \times \dot{\vec{v}})\,,\\
\vec{H}&=&\frac{1}{ {R}_0}\hat{\vec{R}}_0\times \vec{E}\,.
\eeq
Using Eq.~\eqref{CMacc}, we find
\be
\vec{E}_{\rm LW}
=\frac{(q_1+q_2)^2 E_\Omega}{(m_1+m_2) \,c^2  {R}_0 }\hat{\vec{R}}_0\times (\hat{\vec{R}}_0 \times \vec{\mathcal{E}}_{\Omega})\label{ECM1p}\,.
\ee
Finally, in the observer frame, one finds
\beq
\vec{E}_{\rm LW}&=&\frac{(q_1+q_2)^2 E_{\Omega} c_{\Omega t}}{(m_1+m_2) \,c^2 {R}_0 }\bigg[(-c_{\gamma} s^2_{\delta} c^2_{\xi }+s_{\gamma} s_{\delta} c_{\delta} c^2_{\xi}-c_{\gamma} s^2_{\xi}){\mathbf{P}}\nonumber\\
&+&(s_{\gamma} \left(- c^2_{\delta}\right) c^2_{\xi}+c_{\gamma} s_{\delta} c_{\delta} c^2_{\xi}-s_{\gamma} s^2_{\xi}){\mathbf{Q}}\nonumber\\
&+&(c_{\gamma} c_{\delta} s_{\xi} c_{\xi}+s_{\gamma} s_{\delta} s_{\xi} c_{\xi}){\mathbf{N}}\bigg]\,.
\eeq
As we can see from Eq.~\eqref{ECM1p}, this term is due to a non zero acceleration of the CM, that, naturally, depends on the wave perturbation. For two particles with opposite charge, the CM radiation is zero because of $q_1+q_2=0$, and the electric field will be only the one produced by the perturbed dipole.
\section{Angle-action variables}
\label{angleaction}
\subsection{Newtonian dynamics in the Delaunay variables}
\label{angleaction1}
From the reduced Newtonian Lagrangian in the center-of-mass coordinates, using spherical coordinates ($r,\theta,\varphi$),
\begin{equation}
\tilde{L}\equiv\frac{L}{m\nu} = \frac{G m}{r} + \frac{1}{2}\dot{r}^{2} + \frac{1}{2}r^{2}\left(\dot{\theta}^{2}+\sin^{2}\theta\,\dot{\varphi}^{2}\right) \,,
\end{equation}
we determine the conjugate momenta $p_{x}=\partial\tilde{L}/\partial\dot{x}$,
\begin{equation}
p_{r}=\dot{r}\,,\qquad p_{\theta}=r^{2}\dot{\theta}\,,\qquad p_{\varphi}=r^{2}\sin^2\theta\,\dot{\varphi}\,,
\end{equation}
and, performing a Legendre transformation, the reduced Hamiltonian,
\beq
\mathcal{\tilde{H}}_{0} &\equiv& p_{r}\dot{r}+p_{\theta}\dot{\theta}+p_{\varphi}\dot{\varphi}-\tilde{L}\nonumber\\
& = &-\frac{G m}{r} + \frac{1}{2}p_{r}^{2} + \frac{1}{2r^{2}}p_{\theta}^{2} + \frac{1}{2r^{2}\sin^2\theta}p_{\varphi}^{2} \,.
\eeq
The angular momentum $\mathbf{L} = \mathbf{r}\wedge\mathbf{v}$ is then given by
\be
L_{r}=0\,,\qquad L_{\theta}= -\frac{p_{\varphi}}{\sin\theta}\,,\qquad L_{\varphi} = p_{\theta}\,.
\ee
We want to go from the canonical set of variables $\left(r,\theta,\varphi,\,p_{r},p_{\theta},p_{\varphi}\right)$ to a set of canonical angle-action variables, by taking into account the symmetries of the system. We use the modified Delaunay variables that are well suited to described the Keplerian two-body problem. The actions are given by,
\begin{align}
\label{JitoOrbitElem}
& J_{3}=\frac{Gm}{\sqrt{-2E}} \,,\quad J_{2}=\frac{Gm}{\sqrt{-2E}}-L \,,\quad J_{1}=L-L_{z} \,.
\end{align}
The Hamiltonian is then simply
\be
\mathcal{\tilde{H}}_{0} = -\frac{G^{2}m^{2}}{2J_{3}^2}\,.
\ee
We can then derive the frequencies $\Omega_{i}=\partial\tilde{H}/\partial J_{i}$,
\be
\Omega_{3}= \frac{G^{2}m^{2}}{J_{3}^{3}}\,,\qquad \Omega_{2} = 0\,,\qquad \Omega_{1} = 0\,.
\ee
The angles $\theta_{i}$, conjugate variables of the action $J_{i}$, are then linear in time,
\be
\theta_{3}=\Omega_{3}(t-t_{0})+(\theta_{3})_{0} \,,\ \ \theta_{2}=(\theta_{2})_{0}\,,\ \ \theta_{1}=(\theta_{1})_{0}\,,
\ee
with $(\theta_{i})_{0}$ the values of the angle variables at time $t_{0}$.
We can also relate the modified Delaunay variables to the orbital elements $a,e,l,\iota,\omega,\psi$, we get 
\beq
J_{3}&=&\sqrt{Gma} \,,\quad J_{2}=\sqrt{Gma}\left(1-\sqrt{1-e^{2}}\right) \,,\nonumber\\
J_{1}&=&\sqrt{Gma(1-e^{2})}\left(1-\cos\iota\right) \,,\\
\theta_{3}&=&l+\omega+\psi \,,\ \, \theta_{2}=-\left(\omega+\psi\right) \,,\ \, \theta_{1}=-\psi \,.\label{JitoOrbitElem2}
\eeq
Note in particular that these variables are well-defined when $e=0$ and $\iota=0$, which will allow us to perform an expansion for small eccentricity.
\subsection{Perturbation theory}
\label{angleaction2}
When the total Hamiltonian is no longer integrable, it is not possible to write it as a function of the actions only.
The perturbed Hamiltonian in the modified Delaunay variables can be written as
\be
\mathcal{\tilde{H}} = \mathcal{\tilde{H}}_{0}\left(\mathbf{J}\right) + \mathcal{\tilde{H}}_{1}\left(\boldsymbol{\theta},\mathbf{J},t\right)\,,
\ee
where $\mathcal{\tilde{H}}_{0}=-G^{2} m^{2}/(2 J_{3}^{2})$ is the Newtonian Hamiltonian previously studied, and $\tilde{\mathcal{H}}_{1}$ is the perturbation, assumed to be small, $\mathcal{O}(H_{ij})\equiv\mathcal{O}(\varepsilon)\ll 1$. We see that the total Hamiltonian now depends on the actions and angles, but also on time. The dependence on time can be removed by introducing a new coordinate $\tau$ and its conjugate variable $\mathcal{T}$, and transforming the time-dependent Hamiltonian $\mathcal{\tilde{H}}$ into a time-independent Hamiltonian $\mathcal{\tilde{\tilde{H}}}$ in the following way,
\be
\mathcal{\tilde{\tilde{H}}} = \Omega\mathcal{T} + \mathcal{\tilde{H}}\left(\boldsymbol{\theta},\mathbf{J},\tau\right)\,. \label{Ham_tau}
\ee
We see that the Hamilton equations for $\tau$ are
\be
\dot{\tau}=\frac{\partial\mathcal{\tilde{\tilde{H}}}}{\partial\mathcal{T}}=\Omega\,, \qquad \dot{\mathcal{T}} = -\frac{\partial\mathcal{\tilde{H}}_{1}}{\partial\tau}\,,
\ee
such that $\tau = \Omega t$. The other equations for the angle-action variables are unchanged, and are now given by,
\beq
\dot{J_{3}} &=& -\frac{\partial\mathcal{\tilde{H}}_{1}}{\partial\theta_{3}} \,,\qquad \dot{J_{2}} = -\frac{\partial\mathcal{\tilde{H}}_{1}}{\partial\theta_{2}} \,,\qquad \dot{J_{1}} = -\frac{\partial\mathcal{\tilde{H}}_{1}}{\partial\theta_{1}} \,,\nonumber\\
\dot{\theta_{3}}&=&\frac{G^{2}m^{2}}{J_{3}^{3}} +\frac{\partial\mathcal{\tilde{H}}_{1}}{\partial J_{3}} \,,\ \, \dot{\theta_{2}} =\frac{\partial\mathcal{\tilde{H}}_{1}}{\partial J_{2}} \,,\ \, \dot{\theta_{1}}= \frac{\partial\mathcal{\tilde{H}}_{1}}{\partial J_{1}} \,.\label{HamiltonEqs}
\eeq
Using the relations~\eqref{JitoOrbitElem} linking the angle-action coordinates $\left(\boldsymbol{\theta},\mathbf{J}\right)$, and the Hamilton equations~\eqref{HamiltonEqs}, we can see that we directly have the variation of the orbit elements $a,e,l,\iota,\omega,\psi$. As the perturbation is small, we can use the unperturbed (Newtonian) results to evaluate $\tilde{\mathcal{H}}_{1}$ and its derivatives. Then by averaging over one (Newtonian) orbit, we get the secular evolution of the binary. However the Hamiltonian $\mathcal{\tilde{\tilde{H}}}$ is quite complicated and so are the equations~\eqref{HamiltonEqs}.

In order to circumvent these technical difficulties we use Hamiltonian perturbation theory to define a new set of canonical angle-action coordinates $\left(\boldsymbol{\theta}_{\mathrm{new}},\,\boldsymbol{J}_{\mathrm{new}}\right)$ such that the Hamiltonian will only depend on the action variables. We call $\left(\boldsymbol{\theta}^{0},\,\boldsymbol{J}^{0}\right)$ the old variables (including $\tau$ and $\mathcal{T}$). We have
\be
\mathcal{\tilde{\tilde{H}}}\left(\boldsymbol{J}_{\mathrm{new}}\right) = \mathcal{H}_{0}\left(\boldsymbol{J}^{0}\right) + \mathcal{\tilde{H}}_{1}\left(\boldsymbol{\theta}^{0},\mathbf{J}^{0}\right)\,.
\ee
We now consider the generating function
\be
\tilde{S}\left(\boldsymbol{\theta}_{\mathrm{new}},\,\boldsymbol{J}^{0}\right) = \boldsymbol{\theta}_{\mathrm{new}}\cdot\boldsymbol{J}^{0} +\tilde{s}\left(\boldsymbol{\theta}_{\mathrm{new}},\,\boldsymbol{J}^{0}\right)\,,
\ee
where $\tilde{s}=\mathcal{O}\left(\varepsilon\right)$. Then we can rewrite the Hamiltonian, up to $\mathcal{O}\left(\varepsilon^{2}\right)$, as
\be
\mathcal{\tilde{\tilde{H}}}\left(\boldsymbol{J}_{\mathrm{new}}\right) = \mathcal{H}_{0}\left(\boldsymbol{J}_{\mathrm{new}}\right) -\boldsymbol{\Omega}^{0}\cdot\frac{\partial\tilde{s}}{\partial\boldsymbol{\theta}_{\mathrm{new}}} + \mathcal{\tilde{H}}_{1}\left(\boldsymbol{\theta}_{\mathrm{new}},\mathbf{J}_{\mathrm{new}}\right)\,.\nonumber
\ee
where $\boldsymbol{\Omega}^{0}\equiv\frac{\partial\mathcal{H}_{0}}{\partial\mathbf{J}^{0}}$. Now we expand both the perturbed Hamiltonian $\mathcal{\tilde{H}}_{1}$ and $\tilde{s}$ in Fourier series,
\begin{align}
\mathcal{\tilde{H}}_{1}\left(\boldsymbol{\theta}_{\mathrm{new}},\boldsymbol{J}_{\mathrm{new}}\right) & = \sum_{\mathbf{k}}h_{\mathbf{k}}\left(\mathbf{J}_{\mathrm{new}}\right)\mathrm{e}^{i\mathbf{k}\cdot\boldsymbol{\theta}_{\mathrm{new}}} \,,\\
\tilde{s}\left(\boldsymbol{\theta}_{\mathrm{new}},\boldsymbol{J}_{\mathrm{new}}\right) & = i\sum_{\mathbf{k}}s_{\mathbf{k}}\left(\mathbf{J}_{\mathrm{new}}\right)\mathrm{e}^{i\mathbf{k}\cdot\boldsymbol{\theta}_{\mathrm{new}}} \,.
\end{align}
Then the Hamiltonian becomes, up to $\mathcal{O}\left(\varepsilon^{2}\right)$,
\beq\label{Htt}
\mathcal{\tilde{\tilde{H}}}\left(\boldsymbol{J}_{\mathrm{new}}\right) &=& \mathcal{H}_{0}\left(\boldsymbol{J}_{\mathrm{new}}\right) +h_{\mathbf{0}}\left(\boldsymbol{J}_{\mathrm{new}}\right) \nonumber\\
&&\hspace{-.6cm}+\sum_{\mathbf{k}\neq\mathbf{0}}\left[h_{\mathbf{k}}\left(\mathbf{J}_{\mathrm{new}}\right) +\mathbf{k}\cdot\boldsymbol{\Omega}^{0}\left(\mathbf{J}_{\mathrm{new}}\right)s_{\mathbf{k}}\left(\mathbf{J}\right)\right]\mathrm{e}^{i\mathbf{k}\cdot\boldsymbol{\theta}_{\mathrm{new}}} \,.\nonumber
\eeq
As the l.h.s. of Eq.~\eqref{Htt} depends only on the action variables $\mathbf{J}_{\mathrm{new}}$, the r.h.s. should also depends only on this variables. This gives the Fourier coefficients of the generating functions,
\be \label{GeneratingFunction}
s_{\mathbf{k}}\left(\mathbf{J}\right) = - \frac{h_{\mathbf{k}}\left(\mathbf{J}\right)}{\mathbf{k}\cdot\boldsymbol{\Omega}^{0}\left(\mathbf{J}\right)} \qquad \text{for k}\neq 0\,.
\ee
This transformation is valid only when $\mathbf{k}\cdot\boldsymbol{\Omega}^{0}\left(\mathbf{J}\right) \neq 0$. The case
\be
\mathbf{k}\cdot\boldsymbol{\Omega}^{0}\left(\mathbf{J}\right) = 0 \,,
\ee
is called the \textit{problem of small divisors} and it describes the appearance of a resonance at the corresponding frequency. In that case the formalism we are using to describe the binary dynamics is no more valid.

We now consider the coordinate transformation, defined by~\eqref{GeneratingFunction}, to a new set of canonical variables $\left(\boldsymbol{\theta}',\,\boldsymbol{J}'\right)$. The Hamiltonian obtained after this transformation is
\be
\tilde{\mathcal{H}}'\left(\mathbf{J}'\right) \equiv \tilde{\mathcal{H}}_{0}\left(\mathbf{J}'\right) +h_{0}\left(\mathbf{J}'\right) \,.
\ee
It describes the dynamics of the system up to first order included. The new variables are related to the old ones by the relations,
\begin{align}
\mathbf{J}' &= \mathbf{J} +\sum_{\mathbf{k}}\frac{h_{\mathbf{k}}\left(\mathbf{J}\right)}{\mathbf{k}\cdot\boldsymbol{\Omega}^{0}\left(\mathbf{J}\right)}\mathbf{k}\mathrm{e}^{i\mathbf{k}\cdot\boldsymbol{\theta}'} \,,\\
\boldsymbol{\theta}' &= \boldsymbol{\theta} + i\sum_{\mathbf{k}}\frac{\partial}{\partial\mathbf{J}}\left(\frac{h_{\mathbf{k}}\left(\mathbf{J}\right)}{\mathbf{k}\cdot\boldsymbol{\Omega}^{0}\left(\mathbf{J}\right)}\right)\mathrm{e}^{i\mathbf{k}\cdot\boldsymbol{\theta}'} \,.
\end{align}
Finally the dynamics of the system is governed by Hamilton's equation
\begin{align}
\dot{\mathbf{J}'} & = -\frac{\partial\tilde{\mathcal{H}}'}{\partial\boldsymbol{\theta}'} = 0 \,,\\
\dot{\boldsymbol{\theta}'} & = \frac{\partial\tilde{\mathcal{H}}'}{\partial\boldsymbol{J}'} = \frac{\partial\tilde{\mathcal{H}}_{0}}{\partial\boldsymbol{J}'} + \frac{\partial h_{0}}{\partial\boldsymbol{J}'} \,.
\end{align}

\section{The polarizations of the scattered gravitational waves} \label{app:polarization}

\subsection{Parallel to the orbital plane}

Similarly to the usual orbital frequency parameter $x= \left(\frac{Gm\omega_{0}}{c^{3}}\right)^{2/3}$, we define the parameter $X\equiv \left(\frac{Gm\Omega}{c^{3}}\right)^{2/3}$ related to the incoming frequency. When the wave is incident parallel to the orbital plane ($\alpha=0$, $\beta=\pi/2$, $\kappa=\pi/2+\iota$) we find
\begin{widetext}

\begin{equation}\label{hplusform}
\begin{split}
h_{+} = \frac{G \nu m x}{c^2 R} \Biggl\{& -(\cos (2\iota)+3) \cos(2\zeta) + H_{\times} \Biggl[\frac{X^{3/2}}{x^{3/2}} \cos(\Omega t) \Biggl(5 \sin^3\iota \cos\iota \cos(2\zeta)+\frac{1}{48} \biggl(6 \sin(2\iota) (7 \cos(4\zeta)-25) \\
& \quad +\sin (4\iota) (7 \cos(4\zeta)-17)\biggr)\Biggr) +\sin(\Omega t) \Biggl(\frac{X^3}{x^3} \biggl(\frac{1}{6} \sin\iota \cos\iota (\cos(2\iota)+3) \sin(4\zeta)-2 \sin^3\iota \cos\iota \sin(2\zeta)\biggr) \\
& \quad +\frac{1}{4}\sin\iota \cos\iota \biggl(12 \sin^2\iota \sin(2\zeta)+\bigl(8 \cos\iota-3\cos(2\iota)-1\bigr) \sin(4\zeta)\biggr)\Biggr)\Biggr] \\
& + H_{+} \Biggl[\cos (\Omega t) \Biggl(\frac{X^3}{2688 x^3} \biggl(2 \Bigl(661 \cos(2\iota)+18 \cos(4\iota)-21 (\cos(6\iota)+74)\Bigr) \cos(2\zeta) \\
& \quad -7 \Bigl(-65 \cos(2\iota)+6\cos(4\iota)+\cos (6\iota)-198\Bigr) \cos (4\zeta)-448(\cos(2\iota)+3) \sin\zeta +448 (\cos(2\iota)+3) \sin(3\zeta) \\
& \quad +647 \cos(2\iota)+6 \cos(4\iota)-7 \cos(6\iota)-6\biggr) +\frac{1}{2688} \biggl(7 \Bigl(67 \cos(2\iota)-3 \bigl(6 \cos(4\iota) \\
& \quad +\cos(6\iota)-70\bigr)\Bigr) \cos^2(2\zeta)+672 \sin^2\left(\frac{\iota}{2}\right) (\cos(2\iota)-3) \Bigl(\cos\iota \bigl((\cos (2\iota)+3) \cos(4\zeta)-4\bigr)-2 \sin ^2\iota\Bigr) \\
& \quad -\cos(2\zeta) \Bigl(448 (\cos(2\iota)+3) \sin\zeta+3931 \cos(2\iota)+38 \cos(4\iota)-91 \cos(6\iota)-4774\Bigr)\biggr)\Biggr) \\
& \quad +\frac{X^{3/2}}{1344 x^{3/2}} \sin (\Omega t) \Biggl(-6 \sin ^2\iota \biggl(8\cos(2\iota) +35 \cos(4\iota)-619\biggr) \sin(2\zeta) \\
& \quad
-49 (\cos (2\iota)-3) (\cos(2\iota)+3)^2 \sin(4\zeta)\Biggr)\Biggr] \Biggr\}\,,
\end{split}
\end{equation}
and
\begin{equation}\label{hcrossform}
\begin{split}
h_{\times} = \frac{G \nu m x}{c^2 R} \Biggl\{& -4 \cos\iota \sin(2\zeta) + H_{\times} \Biggl[\frac{X^{3/2}}{x^{3/2}} \cos(\Omega t) \Biggl(\frac{7}{3} \sin\iota \cos^2\iota \sin(4\zeta)-\frac{1}{7} \sin^3\iota \sin(2\zeta)\Biggr) \\
& \quad +\sin(\Omega t) \Biggl(\frac{X^3}{21 x^3} \biggl(2 \sin(3\iota)-2 \sin\iota \Bigl(7 \cos^2\iota \cos(4\zeta)+3 \sin^2\iota \cos(2\zeta)+5\Bigr)\biggr) + 8 \sin\iota \cos^2\iota \sin^4\zeta \\
& \quad-\frac{8}{7} \cos(2\zeta) \biggl(\sin^3\iota-7 \sin\iota \cos\iota \sin^2\zeta\biggr)\Biggr)\Biggr] \\
& + H_{+} \Biggl[\frac{X^{3/2}}{192 x^{3/2}} \Biggl(2 \cos\iota \biggl(7 (\cos(4\iota)-17) \cos(4\zeta)-409\biggr)+65 \cos(3\iota)+17 \cos(5\iota)\Biggr) \sin(\Omega t) \\
& \quad +\cos(\Omega t) \Biggl(\frac{1}{48} \sin(2\zeta) \biggl(-24 \sin^2\left(\frac{\iota}{2}\right) (\cos(4\iota)-17) \sin^2\zeta+\cos\iota \Bigl((35-3 \cos(4\iota)) \cos(2\zeta) \\
& \quad -23 \cos(2\iota)+\cos(4\iota)-32 \sin\zeta+38\Bigr)-3 \sin\iota \Bigl(\cos(2\iota)+2 \cos (4\iota)-17\Bigr) \tan\iota\biggr) \\
& \quad -\frac{X^3}{24 x^3} \cos\iota \sin (2\zeta) \biggl((\cos (4\iota)-33) \cos(2\zeta)-32 \sin\zeta+16\biggr)\Biggr)\Biggr] \Biggr\} \,.
\end{split}
\end{equation}

\end{widetext}

\newpage\subsection{Perpendicular to the orbital plane}
For perpendicular incidence ($\alpha=0$, $\beta=\pi/2$, $\kappa=\iota$), we have
\begin{widetext}

\begin{equation}\label{hplusform2}
\begin{split}
h_{+} = \frac{G \nu m x}{c^2 R} \Biggl\{& -(\cos (2\iota)+3) \cos(2\zeta) + H_{\times} \Biggl[\frac{X^{3/2}}{12 x^{3/2}} \cos^2\iota \cos(\Omega t) \Biggl(-7 (\cos(2\iota)+3) \cos(4\zeta)-60 \sin^2\iota \cos(2\zeta) \\
& \quad +17 \cos(2\iota)+75\Biggr) +\sin(\Omega t) \Biggl(\frac{X^3}{6 x^3} \cos^2\iota \biggl(24 \sin^2\iota \sin\zeta \cos\zeta-(\cos (2\iota)+3) \sin(4\zeta)\biggr) \\
& \quad + \cos^2\iota \biggl(-3 \sin^2\iota \sin(2\zeta)-\frac{1}{4} \Bigl(-8 \cos\iota+5 \cos(2\iota)+7\Bigr) \sin(4\zeta)\biggr)\Biggr)\Biggr] \\
& + H_{+} \Biggl[\frac{X^{3/2}}{x^{3/2}} \sin(\Omega t) \Biggl(\frac{1}{112} \sin^2\iota (35 \cos(2\iota)+109) (\cos(2\iota)+3) \sin(2\zeta)+\frac{7}{192} (\cos (2\iota)+3)^3 \sin(4\zeta)\Biggr) \\
& \quad +\cos(\Omega t) \Biggl(\frac{X^3}{2688 x^3} \biggl(-448 (\cos(2\iota)+3) \cos\zeta-448 (\cos(2\iota)+3) \cos(3\zeta)+14 (\cos(2\iota)+3) \Bigl(12 \cos(2\iota) \\
& \quad +\cos (4\iota)+51\Bigr) \cos(4\zeta)-24 \sin^2\iota \Bigl(92 \cos(2\iota)+7 \cos(4\iota)+157\Bigr) \cos(2\zeta)-448 (\cos(2\iota)+3) \sin\zeta \\
& \quad +448 (\cos(2\iota)+3) \sin(3\zeta)+457 \cos(2\iota)+78 \cos(4\iota)+7 \cos(6\iota)+738\biggr) \\
& \quad +\frac{1}{5376} \biggl(448 (\cos(2\iota)+3) \cos\zeta +448 (\cos(2\iota)+3) \cos(3\zeta)+14 (\cos(2\iota)+3) \Bigl(-168 \cos\iota+132 \cos(2\iota) \\
& \quad -24 \cos(3\iota)+15 \cos(4\iota)+109\Bigr) \cos(4\zeta)+16 \sin^2\iota (\cos(2\iota)+3) (91\cos(2\iota)+345) \cos(2\zeta) \\
& \quad +448 (\cos(2\iota)+3) \sin\zeta-448 (\cos(2\iota)+3) \sin(3\zeta)+8400 \cos\iota-4837 \cos(2\iota) \\
& \quad +21 \Bigl(104 \cos(3\iota)-30 \cos(4\iota)+8 \cos(5\iota)+\cos(6\iota)-82\Bigr)\biggr)\Biggr)\Biggr] \Biggr\}\,,
\end{split}
\end{equation}
and
\begin{equation}\label{hcrossform2}
\begin{split}
h_{\times} = \frac{G \nu m x}{c^2 R} \Biggl\{& -4 \cos\iota \sin(2\zeta) + H_{\times} \Biggl[\frac{X^{3/2}}{x^{3/2}} \cos(\Omega t) \Biggl(\frac{1}{7} \sin^2\iota \cos\iota \sin(2\zeta)-\frac{7}{3} \cos^3\iota \sin(4\zeta)\Biggr) \\
& \quad +\sin(\Omega t) \Biggl(\frac{2 X^3}{21 x^3} \cos\iota \biggl(7 \cos^2\iota \cos(4\zeta)+3 \sin^2\iota \cos(2\zeta)-2 \cos(2\iota)+4\biggr) \\
& \quad + \biggl(\frac{8}{7} \cos\iota \cos(2\zeta) \Bigl(14 \sin^2\left(\frac{\iota}{2}\right) \cos\iota \sin^2\zeta+\sin^2\iota\Bigr)-2 \cos^3\iota \sin^2(2\zeta)\biggr)\Biggr)\Biggr] \\
& + H_{+} \Biggl[\frac{X^{3/2}}{192 x^{3/2}} \Biggl(-2 \cos\iota \biggl(14 (\cos(2\iota)+3)^2 \cos(4\zeta)+593\biggr)-269 \cos(3\iota)-17 \cos(5\iota)\Biggr) \sin(\Omega t) \\
& \quad +\cos(\Omega t) \Biggl(\frac{X^3}{192 x^3} \csc\iota \biggl(\Bigl(101 \sin(2\iota)+12 \sin(4\iota)+\sin(6\iota)\Bigr) \sin(4\zeta)-64 \sin(2\iota) \Bigl(\sin\zeta+\sin(3\zeta) \\
& \quad -\cos\zeta+\cos(3\zeta)\Bigr)\biggr) +\frac{1}{192} \biggl(3 \Bigl(-48 \cos(2\iota)+39 \cos(3\iota)-4 \cos(4\iota)+3 \cos(5\iota)-76\Bigr) \sin(4\zeta) \\
& \quad +\cos\iota \Bigl(386 \sin(4\zeta)-64 \cos\zeta+64 \bigl(\sin\zeta+\sin(3\zeta)+\cos(3\zeta)\bigr)\Bigr)+4 \sin^2\left(\frac{\iota}{2}\right) \Bigl(538 \cos\iota \\
& \quad +136 \cos(2\iota)+153 \cos(3\iota)+14 \cos(4\iota)+13 \cos(5\iota)+170\Bigr) \sec\iota \sin(2\zeta)\biggr)\Biggr)\Biggr] \Biggr\} \,.
\end{split}
\end{equation}

\end{widetext}

\newpage
\bibliographystyle{apsrev4}
\bibliography{Ref}

\begin{thebibliography}{49}%
\makeatletter
\providecommand \@ifxundefined [1]{%
 \@ifx{#1\undefined}
}%
\providecommand \@ifnum [1]{%
 \ifnum #1\expandafter \@firstoftwo
 \else \expandafter \@secondoftwo
 \fi
}%
\providecommand \@ifx [1]{%
 \ifx #1\expandafter \@firstoftwo
 \else \expandafter \@secondoftwo
 \fi
}%
\providecommand \natexlab [1]{#1}%
\providecommand \enquote  [1]{``#1''}%
\providecommand \bibnamefont  [1]{#1}%
\providecommand \bibfnamefont [1]{#1}%
\providecommand \citenamefont [1]{#1}%
\providecommand \href@noop [0]{\@secondoftwo}%
\providecommand \href [0]{\begingroup \@sanitize@url \@href}%
\providecommand \@href[1]{\@@startlink{#1}\@@href}%
\providecommand \@@href[1]{\endgroup#1\@@endlink}%
\providecommand \@sanitize@url [0]{\catcode `\\12\catcode `\$12\catcode
  `\&12\catcode `\#12\catcode `\^12\catcode `\_12\catcode `\%12\relax}%
\providecommand \@@startlink[1]{}%
\providecommand \@@endlink[0]{}%
\providecommand \url  [0]{\begingroup\@sanitize@url \@url }%
\providecommand \@url [1]{\endgroup\@href {#1}{\urlprefix }}%
\providecommand \urlprefix  [0]{URL }%
\providecommand \Eprint [0]{\href }%
\providecommand \doibase [0]{http://dx.doi.org/}%
\providecommand \selectlanguage [0]{\@gobble}%
\providecommand \bibinfo  [0]{\@secondoftwo}%
\providecommand \bibfield  [0]{\@secondoftwo}%
\providecommand \translation [1]{[#1]}%
\providecommand \BibitemOpen [0]{}%
\providecommand \bibitemStop [0]{}%
\providecommand \bibitemNoStop [0]{.\EOS\space}%
\providecommand \EOS [0]{\spacefactor3000\relax}%
\providecommand \BibitemShut  [1]{\csname bibitem#1\endcsname}%
\let\auto@bib@innerbib\@empty
\bibitem [{\citenamefont {Abbott} \emph
  {et~al.}(2016{\natexlab{a}})}]{Abbott:2016blz}%
  \BibitemOpen
  \bibfield  {author} {\bibinfo {author} {\bibfnamefont {B.P.} \bibnamefont
  {Abbott}} \emph {et~al.} (\bibinfo {collaboration} {Virgo, LIGO Scientific}),
  }\href {\doibase 10.1103/PhysRevLett.116.061102} {\bibfield  {journal}
  {\bibinfo  {journal} {\emph {Phys. Rev. Lett.}} }\textbf {\bibinfo {volume}
  {116}}, \bibinfo {pages} {061102} (\bibinfo {year} {2016}{\natexlab{a}})},
  \Eprint {http://arxiv.org/abs/1602.03837} {arXiv:1602.03837}\BibitemShut
  {NoStop}%
\bibitem [{\citenamefont {Barack} \emph {et~al.}(2018)}]{Barack:2018yly}%
  \BibitemOpen
  \bibfield  {author} {\bibinfo {author} {\bibfnamefont {L.}~\bibnamefont
  {Barack}} \emph {et~al.}, }\href@noop {} {  (\bibinfo {year} {2018})},
  \Eprint {http://arxiv.org/abs/1806.05195} {arXiv:1806.05195}\BibitemShut
  {NoStop}%
\bibitem [{\citenamefont {Abbott} \emph
  {et~al.}(2016{\natexlab{b}})}]{TheLIGOScientific:2016src}%
  \BibitemOpen
  \bibfield  {author} {\bibinfo {author} {\bibfnamefont {B.P.} \bibnamefont
  {Abbott}} \emph {et~al.} (\bibinfo {collaboration} {Virgo, LIGO Scientific}),
  }\href {\doibase 10.1103/PhysRevLett.116.221101} {\bibfield  {journal}
  {\bibinfo  {journal} {\emph {Phys. Rev. Lett.}} }\textbf {\bibinfo {volume}
  {116}}, \bibinfo {pages} {221101} (\bibinfo {year} {2016}{\natexlab{b}})},
  \Eprint {http://arxiv.org/abs/1602.03841} {arXiv:1602.03841}\BibitemShut
  {NoStop}%
\bibitem [{\citenamefont {Yunes} \emph {et~al.}(2016)\citenamefont {Yunes},
  \citenamefont {Yagi}, and \citenamefont {Pretorius}}]{Yunes:2016jcc}%
  \BibitemOpen
  \bibfield  {author} {\bibinfo {author} {\bibfnamefont {N.}~\bibnamefont
  {Yunes}}, \bibinfo {author} {\bibfnamefont {K.}~\bibnamefont {Yagi}},  and
  \bibinfo {author} {\bibfnamefont {F.}~\bibnamefont {Pretorius}}, }\href
  {\doibase 10.1103/PhysRevD.94.084002} {\bibfield  {journal} {\bibinfo
  {journal} {\emph {Phys. Rev.}} }\textbf {\bibinfo {volume} {D94}}, \bibinfo
  {pages} {084002} (\bibinfo {year} {2016})}, \Eprint
  {http://arxiv.org/abs/1603.08955} {arXiv:1603.08955}\BibitemShut {NoStop}%
\bibitem [{\citenamefont {Berti} \emph {et~al.}(2005)\citenamefont {Berti},
  \citenamefont {Buonanno}, and \citenamefont {Will}}]{Berti:2004bd}%
  \BibitemOpen
  \bibfield  {author} {\bibinfo {author} {\bibfnamefont {E.}~\bibnamefont
  {Berti}}, \bibinfo {author} {\bibfnamefont {A.}~\bibnamefont {Buonanno}},
  and \bibinfo {author} {\bibfnamefont {C.M.} \bibnamefont {Will}}, }\href
  {\doibase 10.1103/PhysRevD.71.084025} {\bibfield  {journal} {\bibinfo
  {journal} {\emph {Phys. Rev.}} }\textbf {\bibinfo {volume} {D71}}, \bibinfo
  {pages} {084025} (\bibinfo {year} {2005})}, \Eprint
  {http://arxiv.org/abs/gr-qc/0411129} {arXiv:gr-qc/0411129}\BibitemShut
  {NoStop}%
\bibitem [{\citenamefont {Amaro-Seoane} \emph {et~al.}(2007)\citenamefont
  {Amaro-Seoane}, \citenamefont {Gair}, \citenamefont {Freitag}, \citenamefont
  {Coleman~Miller}, \citenamefont {Mandel}, \citenamefont {Cutler}, and
  \citenamefont {Babak}}]{AmaroSeoane:2007aw}%
  \BibitemOpen
  \bibfield  {author} {\bibinfo {author} {\bibfnamefont {P.}~\bibnamefont
  {Amaro-Seoane}}, \bibinfo {author} {\bibfnamefont {J.R.} \bibnamefont
  {Gair}}, \bibinfo {author} {\bibfnamefont {M.}~\bibnamefont {Freitag}},
  \bibinfo {author} {\bibfnamefont {M.}~\bibnamefont {Coleman~Miller}},
  \bibinfo {author} {\bibfnamefont {I.}~\bibnamefont {Mandel}}, \bibinfo
  {author} {\bibfnamefont {C.J.} \bibnamefont {Cutler}},  and \bibinfo {author}
  {\bibfnamefont {S.}~\bibnamefont {Babak}}, }\href {\doibase
  10.1088/0264-9381/24/17/R01} {\bibfield  {journal} {\bibinfo  {journal}
  {\emph {Class. Quant. Grav.}} }\textbf {\bibinfo {volume} {24}}, \bibinfo
  {pages} {R113} (\bibinfo {year} {2007})}, \Eprint
  {http://arxiv.org/abs/astro-ph/0703495} {arXiv:astro-ph/0703495}\BibitemShut
  {NoStop}%
\bibitem [{\citenamefont {Barausse} \emph {et~al.}(2016)\citenamefont
  {Barausse}, \citenamefont {Yunes}, and \citenamefont
  {Chamberlain}}]{Barausse:2016eii}%
  \BibitemOpen
  \bibfield  {author} {\bibinfo {author} {\bibfnamefont {E.}~\bibnamefont
  {Barausse}}, \bibinfo {author} {\bibfnamefont {N.}~\bibnamefont {Yunes}},
  and \bibinfo {author} {\bibfnamefont {K.}~\bibnamefont {Chamberlain}}, }\href
  {\doibase 10.1103/PhysRevLett.116.241104} {\bibfield  {journal} {\bibinfo
  {journal} {\emph {Phys. Rev. Lett.}} }\textbf {\bibinfo {volume} {116}},
  \bibinfo {pages} {241104} (\bibinfo {year} {2016})}, \Eprint
  {http://arxiv.org/abs/1603.04075} {arXiv:1603.04075}\BibitemShut {NoStop}%
\bibitem [{\citenamefont {Cardoso} \emph
  {et~al.}(2016{\natexlab{a}})\citenamefont {Cardoso}, \citenamefont {Macedo},
  \citenamefont {Pani}, and \citenamefont {Ferrari}}]{Cardoso:2016olt}%
  \BibitemOpen
  \bibfield  {author} {\bibinfo {author} {\bibfnamefont {V.}~\bibnamefont
  {Cardoso}}, \bibinfo {author} {\bibfnamefont {C.F.B.} \bibnamefont {Macedo}},
  \bibinfo {author} {\bibfnamefont {P.}~\bibnamefont {Pani}},  and \bibinfo
  {author} {\bibfnamefont {V.}~\bibnamefont {Ferrari}}, }\href {\doibase
  10.1088/1475-7516/2016/05/054} {\bibfield  {journal} {\bibinfo  {journal}
  {\emph {JCAP}} }\textbf {\bibinfo {volume} {1605}}, \bibinfo {pages} {054}
  (\bibinfo {year} {2016}{\natexlab{a}})}, \Eprint
  {http://arxiv.org/abs/1604.07845} {arXiv:1604.07845}\BibitemShut {NoStop}%
\bibitem [{\citenamefont {Arvanitaki} \emph {et~al.}(2017)\citenamefont
  {Arvanitaki}, \citenamefont {Baryakhtar}, \citenamefont {Dimopoulos},
  \citenamefont {Dubovsky}, and \citenamefont {Lasenby}}]{Arvanitaki:2016qwi}%
  \BibitemOpen
  \bibfield  {author} {\bibinfo {author} {\bibfnamefont {A.}~\bibnamefont
  {Arvanitaki}}, \bibinfo {author} {\bibfnamefont {M.}~\bibnamefont
  {Baryakhtar}}, \bibinfo {author} {\bibfnamefont {S.}~\bibnamefont
  {Dimopoulos}}, \bibinfo {author} {\bibfnamefont {S.}~\bibnamefont
  {Dubovsky}},  and \bibinfo {author} {\bibfnamefont {R.}~\bibnamefont
  {Lasenby}}, }\href {\doibase 10.1103/PhysRevD.95.043001} {\bibfield
  {journal} {\bibinfo  {journal} {\emph {Phys. Rev.}} }\textbf {\bibinfo
  {volume} {D95}}, \bibinfo {pages} {043001} (\bibinfo {year} {2017})}, \Eprint
  {http://arxiv.org/abs/1604.03958} {arXiv:1604.03958}\BibitemShut {NoStop}%
\bibitem [{\citenamefont {Brito} \emph {et~al.}(2017)\citenamefont {Brito},
  \citenamefont {Ghosh}, \citenamefont {Barausse}, \citenamefont {Berti},
  \citenamefont {Cardoso}, \citenamefont {Dvorkin}, \citenamefont {Klein}, and
  \citenamefont {Pani}}]{Brito:2017zvb}%
  \BibitemOpen
  \bibfield  {author} {\bibinfo {author} {\bibfnamefont {R.}~\bibnamefont
  {Brito}}, \bibinfo {author} {\bibfnamefont {S.}~\bibnamefont {Ghosh}},
  \bibinfo {author} {\bibfnamefont {E.}~\bibnamefont {Barausse}}, \bibinfo
  {author} {\bibfnamefont {E.}~\bibnamefont {Berti}}, \bibinfo {author}
  {\bibfnamefont {V.}~\bibnamefont {Cardoso}}, \bibinfo {author} {\bibfnamefont
  {I.}~\bibnamefont {Dvorkin}}, \bibinfo {author} {\bibfnamefont
  {A.}~\bibnamefont {Klein}},  and \bibinfo {author} {\bibfnamefont
  {P.}~\bibnamefont {Pani}}, }\href {\doibase 10.1103/PhysRevD.96.064050}
  {\bibfield  {journal} {\bibinfo  {journal} {\emph {Phys. Rev.}} }\textbf
  {\bibinfo {volume} {D96}}, \bibinfo {pages} {064050} (\bibinfo {year}
  {2017})}, \Eprint {http://arxiv.org/abs/1706.06311}
  {arXiv:1706.06311}\BibitemShut {NoStop}%
\bibitem [{\citenamefont {Eda} \emph {et~al.}(2013)\citenamefont {Eda},
  \citenamefont {Itoh}, \citenamefont {Kuroyanagi}, and \citenamefont
  {Silk}}]{Eda:2013gg}%
  \BibitemOpen
  \bibfield  {author} {\bibinfo {author} {\bibfnamefont {K.}~\bibnamefont
  {Eda}}, \bibinfo {author} {\bibfnamefont {Y.}~\bibnamefont {Itoh}}, \bibinfo
  {author} {\bibfnamefont {S.}~\bibnamefont {Kuroyanagi}},  and \bibinfo
  {author} {\bibfnamefont {J.}~\bibnamefont {Silk}}, }\href {\doibase
  10.1103/PhysRevLett.110.221101} {\bibfield  {journal} {\bibinfo  {journal}
  {\emph {Phys. Rev. Lett.}} }\textbf {\bibinfo {volume} {110}}, \bibinfo
  {pages} {221101} (\bibinfo {year} {2013})}, \Eprint
  {http://arxiv.org/abs/1301.5971} {arXiv:1301.5971}\BibitemShut {NoStop}%
\bibitem [{\citenamefont {Barausse} \emph {et~al.}(2014)\citenamefont
  {Barausse}, \citenamefont {Cardoso}, and \citenamefont
  {Pani}}]{Barausse:2014tra}%
  \BibitemOpen
  \bibfield  {author} {\bibinfo {author} {\bibfnamefont {E.}~\bibnamefont
  {Barausse}}, \bibinfo {author} {\bibfnamefont {V.}~\bibnamefont {Cardoso}},
  and \bibinfo {author} {\bibfnamefont {P.}~\bibnamefont {Pani}}, }\href
  {\doibase 10.1103/PhysRevD.89.104059} {\bibfield  {journal} {\bibinfo
  {journal} {\emph {Phys. Rev.}} }\textbf {\bibinfo {volume} {D89}}, \bibinfo
  {pages} {104059} (\bibinfo {year} {2014})}, \Eprint
  {http://arxiv.org/abs/1404.7149} {arXiv:1404.7149}\BibitemShut {NoStop}%
\bibitem [{\citenamefont {Barausse} \emph {et~al.}(2015)\citenamefont
  {Barausse}, \citenamefont {Cardoso}, and \citenamefont
  {Pani}}]{Barausse:2014pra}%
  \BibitemOpen
  \bibfield  {author} {\bibinfo {author} {\bibfnamefont {E.}~\bibnamefont
  {Barausse}}, \bibinfo {author} {\bibfnamefont {V.}~\bibnamefont {Cardoso}},
  and \bibinfo {author} {\bibfnamefont {P.}~\bibnamefont {Pani}}, }\bibfield
  {booktitle} {\emph {\bibinfo {booktitle} {{Proceedings, 10th International
  LISA Symposium}}}, }\href {\doibase 10.1088/1742-6596/610/1/012044}
  {\bibfield  {journal} {\bibinfo  {journal} {\emph {J. Phys. Conf. Ser.}}
  }\textbf {\bibinfo {volume} {610}}, \bibinfo {pages} {012044} (\bibinfo
  {year} {2015})}, \Eprint {http://arxiv.org/abs/1404.7140}
  {arXiv:1404.7140}\BibitemShut {NoStop}%
\bibitem [{\citenamefont {Cardoso} \emph {et~al.}(2017)\citenamefont {Cardoso},
  \citenamefont {Franzin}, \citenamefont {Maselli}, \citenamefont {Pani}, and
  \citenamefont {Raposo}}]{Cardoso:2017cfl}%
  \BibitemOpen
  \bibfield  {author} {\bibinfo {author} {\bibfnamefont {V.}~\bibnamefont
  {Cardoso}}, \bibinfo {author} {\bibfnamefont {E.}~\bibnamefont {Franzin}},
  \bibinfo {author} {\bibfnamefont {A.}~\bibnamefont {Maselli}}, \bibinfo
  {author} {\bibfnamefont {P.}~\bibnamefont {Pani}},  and \bibinfo {author}
  {\bibfnamefont {G.}~\bibnamefont {Raposo}}, }\href {\doibase
  10.1103/PhysRevD.95.084014} {\bibfield  {journal} {\bibinfo  {journal} {\emph
  {Phys. Rev.}} }\textbf {\bibinfo {volume} {D95}}, \bibinfo {pages} {084014}
  (\bibinfo {year} {2017})}, \Eprint {http://arxiv.org/abs/1701.01116}
  {arXiv:1701.01116}\BibitemShut {NoStop}%
\bibitem [{\citenamefont {Sennett} \emph {et~al.}(2017)\citenamefont {Sennett},
  \citenamefont {Hinderer}, \citenamefont {Steinhoff}, \citenamefont
  {Buonanno}, and \citenamefont {Ossokine}}]{Sennett:2017etc}%
  \BibitemOpen
  \bibfield  {author} {\bibinfo {author} {\bibfnamefont {N.}~\bibnamefont
  {Sennett}}, \bibinfo {author} {\bibfnamefont {T.}~\bibnamefont {Hinderer}},
  \bibinfo {author} {\bibfnamefont {J.}~\bibnamefont {Steinhoff}}, \bibinfo
  {author} {\bibfnamefont {A.}~\bibnamefont {Buonanno}},  and \bibinfo {author}
  {\bibfnamefont {S.}~\bibnamefont {Ossokine}}, }\href {\doibase
  10.1103/PhysRevD.96.024002} {\bibfield  {journal} {\bibinfo  {journal} {\emph
  {Phys. Rev.}} }\textbf {\bibinfo {volume} {D96}}, \bibinfo {pages} {024002}
  (\bibinfo {year} {2017})}, \Eprint {http://arxiv.org/abs/1704.08651}
  {arXiv:1704.08651}\BibitemShut {NoStop}%
\bibitem [{\citenamefont {Cardoso} and \citenamefont
  {Pani}(2017{\natexlab{a}})}]{Cardoso:2017njb}%
  \BibitemOpen
  \bibfield  {author} {\bibinfo {author} {\bibfnamefont {V.}~\bibnamefont
  {Cardoso}} and \bibinfo {author} {\bibfnamefont {P.}~\bibnamefont {Pani}},
  }\href@noop {} {  (\bibinfo {year} {2017}{\natexlab{a}})}, \Eprint
  {http://arxiv.org/abs/1707.03021} {arXiv:1707.03021}\BibitemShut {NoStop}%
\bibitem [{\citenamefont {Cardoso} and \citenamefont
  {Pani}(2017{\natexlab{b}})}]{Cardoso:2017cqb}%
  \BibitemOpen
  \bibfield  {author} {\bibinfo {author} {\bibfnamefont {V.}~\bibnamefont
  {Cardoso}} and \bibinfo {author} {\bibfnamefont {P.}~\bibnamefont {Pani}},
  }\href {\doibase 10.1038/s41550-017-0225-y} {\bibfield  {journal} {\bibinfo
  {journal} {\emph {Nat. Astron.}} }\textbf {\bibinfo {volume} {1}}, \bibinfo
  {pages} {586} (\bibinfo {year} {2017}{\natexlab{b}})}, \Eprint
  {http://arxiv.org/abs/1709.01525} {arXiv:1709.01525}\BibitemShut {NoStop}%
\bibitem [{\citenamefont {Berti} \emph {et~al.}(2006)\citenamefont {Berti},
  \citenamefont {Cardoso}, and \citenamefont {Will}}]{Berti:2005ys}%
  \BibitemOpen
  \bibfield  {author} {\bibinfo {author} {\bibfnamefont {E.}~\bibnamefont
  {Berti}}, \bibinfo {author} {\bibfnamefont {V.}~\bibnamefont {Cardoso}},  and
  \bibinfo {author} {\bibfnamefont {C.M.} \bibnamefont {Will}}, }\href
  {\doibase 10.1103/PhysRevD.73.064030} {\bibfield  {journal} {\bibinfo
  {journal} {\emph {Phys. Rev.}} }\textbf {\bibinfo {volume} {D73}}, \bibinfo
  {pages} {064030} (\bibinfo {year} {2006})}, \Eprint
  {http://arxiv.org/abs/gr-qc/0512160} {arXiv:gr-qc/0512160}\BibitemShut
  {NoStop}%
\bibitem [{\citenamefont {Berti} \emph {et~al.}(2016)\citenamefont {Berti},
  \citenamefont {Sesana}, \citenamefont {Barausse}, \citenamefont {Cardoso},
  and \citenamefont {Belczynski}}]{Berti:2016lat}%
  \BibitemOpen
  \bibfield  {author} {\bibinfo {author} {\bibfnamefont {E.}~\bibnamefont
  {Berti}}, \bibinfo {author} {\bibfnamefont {A.}~\bibnamefont {Sesana}},
  \bibinfo {author} {\bibfnamefont {E.}~\bibnamefont {Barausse}}, \bibinfo
  {author} {\bibfnamefont {V.}~\bibnamefont {Cardoso}},  and \bibinfo {author}
  {\bibfnamefont {K.}~\bibnamefont {Belczynski}}, }\href {\doibase
  10.1103/PhysRevLett.117.101102} {\bibfield  {journal} {\bibinfo  {journal}
  {\emph {Phys. Rev. Lett.}} }\textbf {\bibinfo {volume} {117}}, \bibinfo
  {pages} {101102} (\bibinfo {year} {2016})}, \Eprint
  {http://arxiv.org/abs/1605.09286} {arXiv:1605.09286}\BibitemShut {NoStop}%
\bibitem [{\citenamefont {Yang} \emph {et~al.}(2017)\citenamefont {Yang},
  \citenamefont {Yagi}, \citenamefont {Blackman}, \citenamefont {Lehner},
  \citenamefont {Paschalidis}, \citenamefont {Pretorius}, and \citenamefont
  {Yunes}}]{Yang:2017zxs}%
  \BibitemOpen
  \bibfield  {author} {\bibinfo {author} {\bibfnamefont {H.}~\bibnamefont
  {Yang}}, \bibinfo {author} {\bibfnamefont {K.}~\bibnamefont {Yagi}}, \bibinfo
  {author} {\bibfnamefont {J.}~\bibnamefont {Blackman}}, \bibinfo {author}
  {\bibfnamefont {L.}~\bibnamefont {Lehner}}, \bibinfo {author} {\bibfnamefont
  {V.}~\bibnamefont {Paschalidis}}, \bibinfo {author} {\bibfnamefont
  {F.}~\bibnamefont {Pretorius}},  and \bibinfo {author} {\bibfnamefont
  {N.}~\bibnamefont {Yunes}}, }\href {\doibase 10.1103/PhysRevLett.118.161101}
  {\bibfield  {journal} {\bibinfo  {journal} {\emph {Phys. Rev. Lett.}}
  }\textbf {\bibinfo {volume} {118}}, \bibinfo {pages} {161101} (\bibinfo
  {year} {2017})}, \Eprint {http://arxiv.org/abs/1701.05808}
  {arXiv:1701.05808}\BibitemShut {NoStop}%
\bibitem [{\citenamefont {Cardoso} \emph
  {et~al.}(2016{\natexlab{b}})\citenamefont {Cardoso}, \citenamefont {Franzin},
  and \citenamefont {Pani}}]{Cardoso:2016rao}%
  \BibitemOpen
  \bibfield  {author} {\bibinfo {author} {\bibfnamefont {V.}~\bibnamefont
  {Cardoso}}, \bibinfo {author} {\bibfnamefont {E.}~\bibnamefont {Franzin}},
  and \bibinfo {author} {\bibfnamefont {P.}~\bibnamefont {Pani}}, }\href
  {\doibase 10.1103/PhysRevLett.117.089902, 10.1103/PhysRevLett.116.171101}
  {\bibfield  {journal} {\bibinfo  {journal} {\emph {Phys. Rev. Lett.}}
  }\textbf {\bibinfo {volume} {116}}, \bibinfo {pages} {171101} (\bibinfo
  {year} {2016}{\natexlab{b}})}, \bibinfo {note} {[Erratum: Phys. Rev.
  Lett.117,no.8,089902(2016)]}, \Eprint {http://arxiv.org/abs/1602.07309}
  {arXiv:1602.07309}\BibitemShut {NoStop}%
\bibitem [{\citenamefont {Abbott} \emph
  {et~al.}(2017)}]{TheLIGOScientific:2017qsa}%
  \BibitemOpen
  \bibfield  {author} {\bibinfo {author} {\bibfnamefont {B.}~\bibnamefont
  {Abbott}} \emph {et~al.} (\bibinfo {collaboration} {Virgo, LIGO Scientific}),
  }\href {\doibase 10.1103/PhysRevLett.119.161101} {\bibfield  {journal}
  {\bibinfo  {journal} {\emph {Phys. Rev. Lett.}} }\textbf {\bibinfo {volume}
  {119}}, \bibinfo {pages} {161101} (\bibinfo {year} {2017})}, \Eprint
  {http://arxiv.org/abs/1710.05832} {arXiv:1710.05832}\BibitemShut {NoStop}%
\bibitem [{\citenamefont {Ezquiaga} and \citenamefont
  {Zumalac\'arregui}(2017)}]{Ezquiaga:2017ekz}%
  \BibitemOpen
  \bibfield  {author} {\bibinfo {author} {\bibfnamefont {J.M.} \bibnamefont
  {Ezquiaga}} and \bibinfo {author} {\bibfnamefont {M.}~\bibnamefont
  {Zumalac\'arregui}}, }\href {\doibase 10.1103/PhysRevLett.119.251304}
  {\bibfield  {journal} {\bibinfo  {journal} {\emph {Phys. Rev. Lett.}}
  }\textbf {\bibinfo {volume} {119}}, \bibinfo {pages} {251304} (\bibinfo
  {year} {2017})}, \Eprint {http://arxiv.org/abs/1710.05901}
  {arXiv:1710.05901}\BibitemShut {NoStop}%
\bibitem [{\citenamefont {Creminelli} and \citenamefont
  {Vernizzi}(2017)}]{Creminelli:2017sry}%
  \BibitemOpen
  \bibfield  {author} {\bibinfo {author} {\bibfnamefont {P.}~\bibnamefont
  {Creminelli}} and \bibinfo {author} {\bibfnamefont {F.}~\bibnamefont
  {Vernizzi}}, }\href {\doibase 10.1103/PhysRevLett.119.251302} {\bibfield
  {journal} {\bibinfo  {journal} {\emph {Phys. Rev. Lett.}} }\textbf {\bibinfo
  {volume} {119}}, \bibinfo {pages} {251302} (\bibinfo {year} {2017})}, \Eprint
  {http://arxiv.org/abs/1710.05877} {arXiv:1710.05877}\BibitemShut {NoStop}%
\bibitem [{\citenamefont {Baker} \emph {et~al.}(2017)\citenamefont {Baker},
  \citenamefont {Bellini}, \citenamefont {Ferreira}, \citenamefont {Lagos},
  \citenamefont {Noller}, and \citenamefont {Sawicki}}]{Baker:2017hug}%
  \BibitemOpen
  \bibfield  {author} {\bibinfo {author} {\bibfnamefont {T.}~\bibnamefont
  {Baker}}, \bibinfo {author} {\bibfnamefont {E.}~\bibnamefont {Bellini}},
  \bibinfo {author} {\bibfnamefont {P.G.} \bibnamefont {Ferreira}}, \bibinfo
  {author} {\bibfnamefont {M.}~\bibnamefont {Lagos}}, \bibinfo {author}
  {\bibfnamefont {J.}~\bibnamefont {Noller}},  and \bibinfo {author}
  {\bibfnamefont {I.}~\bibnamefont {Sawicki}}, }\href {\doibase
  10.1103/PhysRevLett.119.251301} {\bibfield  {journal} {\bibinfo  {journal}
  {\emph {Phys. Rev. Lett.}} }\textbf {\bibinfo {volume} {119}}, \bibinfo
  {pages} {251301} (\bibinfo {year} {2017})}, \Eprint
  {http://arxiv.org/abs/1710.06394} {arXiv:1710.06394}\BibitemShut {NoStop}%
\bibitem [{\citenamefont {Grishchuk} and \citenamefont
  {Polnarev}(1981)}]{Grishchuk:1981fp}%
  \BibitemOpen
  \bibfield  {author} {\bibinfo {author} {\bibfnamefont {L.P.} \bibnamefont
  {Grishchuk}} and \bibinfo {author} {\bibfnamefont {A.G.} \bibnamefont
  {Polnarev}}, }\href@noop {} {  (\bibinfo {year} {1981})}\BibitemShut
  {NoStop}%
\bibitem [{\citenamefont {Deruelle} and \citenamefont
  {Piran}(1984)}]{Deruelle:1984hq}%
  \BibitemOpen
  \bibinfo {editor} {\bibfnamefont {N.}~\bibnamefont {Deruelle}} and \bibinfo
  {editor} {\bibfnamefont {T.}~\bibnamefont {Piran}}, eds., \href@noop {}
  {\emph {\bibinfo {title} {{GRAVITATIONAL RADIATION. PROCEEDINGS, SUMMER
  SCHOOL, NATO ADVANCED STUDY INSTITUTE, LES HOUCHES, FRANCE, JUNE 2-21,
  1982}}}} (\bibinfo {year} {1984})\BibitemShut {NoStop}%
\bibitem [{\citenamefont {Baym} \emph {et~al.}(2017)\citenamefont {Baym},
  \citenamefont {Patil}, and \citenamefont {Pethick}}]{Baym:2017xvh}%
  \BibitemOpen
  \bibfield  {author} {\bibinfo {author} {\bibfnamefont {G.}~\bibnamefont
  {Baym}}, \bibinfo {author} {\bibfnamefont {S.P.} \bibnamefont {Patil}},  and
  \bibinfo {author} {\bibfnamefont {C.J.} \bibnamefont {Pethick}}, }\href
  {\doibase 10.1103/PhysRevD.96.084033} {\bibfield  {journal} {\bibinfo
  {journal} {\emph {Phys. Rev.}} }\textbf {\bibinfo {volume} {D96}}, \bibinfo
  {pages} {084033} (\bibinfo {year} {2017})}, \Eprint
  {http://arxiv.org/abs/1707.05192} {arXiv:1707.05192}\BibitemShut {NoStop}%
\bibitem [{\citenamefont {Flauger} and \citenamefont
  {Weinberg}(2017)}]{Flauger:2017ged}%
  \BibitemOpen
  \bibfield  {author} {\bibinfo {author} {\bibfnamefont {R.}~\bibnamefont
  {Flauger}} and \bibinfo {author} {\bibfnamefont {S.}~\bibnamefont
  {Weinberg}}, }\href@noop {} {  (\bibinfo {year} {2017})}, \Eprint
  {http://arxiv.org/abs/1801.00386} {arXiv:1801.00386}\BibitemShut {NoStop}%
\bibitem [{\citenamefont {Dev} \emph {et~al.}(2017)\citenamefont {Dev},
  \citenamefont {Lindner}, and \citenamefont {Ohmer}}]{Dev:2016hxv}%
  \BibitemOpen
  \bibfield  {author} {\bibinfo {author} {\bibfnamefont {P.S.B.} \bibnamefont
  {Dev}}, \bibinfo {author} {\bibfnamefont {M.}~\bibnamefont {Lindner}},  and
  \bibinfo {author} {\bibfnamefont {S.}~\bibnamefont {Ohmer}}, }\href {\doibase
  10.1016/j.physletb.2017.08.043} {\bibfield  {journal} {\bibinfo  {journal}
  {\emph {Phys. Lett.}} }\textbf {\bibinfo {volume} {B773}}, \bibinfo {pages}
  {219} (\bibinfo {year} {2017})}, \Eprint {http://arxiv.org/abs/1609.03939}
  {arXiv:1609.03939}\BibitemShut {NoStop}%
\bibitem [{\citenamefont {Cai} \emph {et~al.}(2018)\citenamefont {Cai},
  \citenamefont {Liu}, and \citenamefont {Wang}}]{Cai:2017buj}%
  \BibitemOpen
  \bibfield  {author} {\bibinfo {author} {\bibfnamefont {R.G.} \bibnamefont
  {Cai}}, \bibinfo {author} {\bibfnamefont {T.B.} \bibnamefont {Liu}},  and
  \bibinfo {author} {\bibfnamefont {S.J.} \bibnamefont {Wang}}, }\href
  {\doibase 10.1103/PhysRevD.97.023027} {\bibfield  {journal} {\bibinfo
  {journal} {\emph {Phys. Rev.}} }\textbf {\bibinfo {volume} {D97}}, \bibinfo
  {pages} {023027} (\bibinfo {year} {2018})}, \Eprint
  {http://arxiv.org/abs/1710.02425} {arXiv:1710.02425}\BibitemShut {NoStop}%
\bibitem [{\citenamefont {Khmelnitsky} and \citenamefont
  {Rubakov}(2014)}]{Khmelnitsky:2013lxt}%
  \BibitemOpen
  \bibfield  {author} {\bibinfo {author} {\bibfnamefont {A.}~\bibnamefont
  {Khmelnitsky}} and \bibinfo {author} {\bibfnamefont {V.}~\bibnamefont
  {Rubakov}}, }\href {\doibase 10.1088/1475-7516/2014/02/019} {\bibfield
  {journal} {\bibinfo  {journal} {\emph {JCAP}} }\textbf {\bibinfo {volume}
  {1402}}, \bibinfo {pages} {019} (\bibinfo {year} {2014})}, \Eprint
  {http://arxiv.org/abs/1309.5888} {arXiv:1309.5888}\BibitemShut {NoStop}%
\bibitem [{\citenamefont {Turner}(1979)}]{Turner:1979yn}%
  \BibitemOpen
  \bibfield  {author} {\bibinfo {author} {\bibfnamefont {M.S.} \bibnamefont
  {Turner}}, }\href {\doibase 10.1086/157429} {\bibfield  {journal} {\bibinfo
  {journal} {\emph {Astrophys. J.}} }\textbf {\bibinfo {volume} {233}},
  \bibinfo {pages} {685} (\bibinfo {year} {1979})}\BibitemShut {NoStop}%
\bibitem [{\citenamefont {{Mashhoon}}(1978)}]{1978ApJ...223..285M}%
  \BibitemOpen
  \bibfield  {author} {\bibinfo {author} {\bibfnamefont {B.}~\bibnamefont
  {{Mashhoon}}}, }\href {\doibase 10.1086/156262} {\bibfield  {journal}
  {\bibinfo  {journal} {\emph {\apj}} }\textbf {\bibinfo {volume} {223}},
  \bibinfo {pages} {285} (\bibinfo {year} {1978})}\BibitemShut {NoStop}%
\bibitem [{\citenamefont {McKernan} \emph {et~al.}(2014)\citenamefont
  {McKernan}, \citenamefont {Ford}, \citenamefont {Kocsis}, and \citenamefont
  {Haiman}}]{McKernan:2014hha}%
  \BibitemOpen
  \bibfield  {author} {\bibinfo {author} {\bibfnamefont {B.}~\bibnamefont
  {McKernan}}, \bibinfo {author} {\bibfnamefont {K.E.S.} \bibnamefont {Ford}},
  \bibinfo {author} {\bibfnamefont {B.}~\bibnamefont {Kocsis}},  and \bibinfo
  {author} {\bibfnamefont {Z.}~\bibnamefont {Haiman}}, }\href {\doibase
  10.1093/mnrasl/slu136} {\bibfield  {journal} {\bibinfo  {journal} {\emph
  {Mon. Not. Roy. Astron. Soc.}} }\textbf {\bibinfo {volume} {445}}, \bibinfo
  {pages} {74} (\bibinfo {year} {2014})}, \Eprint
  {http://arxiv.org/abs/1405.1414} {arXiv:1405.1414}\BibitemShut {NoStop}%
\bibitem [{\citenamefont {Lopes} and \citenamefont
  {Silk}(2014)}]{Lopes:2014dba}%
  \BibitemOpen
  \bibfield  {author} {\bibinfo {author} {\bibfnamefont {I.}~\bibnamefont
  {Lopes}} and \bibinfo {author} {\bibfnamefont {J.}~\bibnamefont {Silk}},
  }\href {\doibase 10.1088/0004-637X/794/1/32} {\bibfield  {journal} {\bibinfo
  {journal} {\emph {Astrophys. J.}} }\textbf {\bibinfo {volume} {794}},
  \bibinfo {pages} {32} (\bibinfo {year} {2014})}, \Eprint
  {http://arxiv.org/abs/1405.0292} {arXiv:1405.0292}\BibitemShut {NoStop}%
\bibitem [{\citenamefont {Hui} \emph {et~al.}(2013)\citenamefont {Hui},
  \citenamefont {McWilliams}, and \citenamefont {Yang}}]{Hui:2012yp}%
  \BibitemOpen
  \bibfield  {author} {\bibinfo {author} {\bibfnamefont {L.}~\bibnamefont
  {Hui}}, \bibinfo {author} {\bibfnamefont {S.T.} \bibnamefont {McWilliams}},
  and \bibinfo {author} {\bibfnamefont {I.S.} \bibnamefont {Yang}}, }\href
  {\doibase 10.1103/PhysRevD.87.084009} {\bibfield  {journal} {\bibinfo
  {journal} {\emph {Phys. Rev.}} }\textbf {\bibinfo {volume} {D87}}, \bibinfo
  {pages} {084009} (\bibinfo {year} {2013})}, \Eprint
  {http://arxiv.org/abs/1212.2623} {arXiv:1212.2623}\BibitemShut {NoStop}%
\bibitem [{\citenamefont {Poisson} and \citenamefont
  {Will}(2014)}]{PoissonWill2014}%
  \BibitemOpen
  \bibfield  {author} {\bibinfo {author} {\bibfnamefont {E.}~\bibnamefont
  {Poisson}} and \bibinfo {author} {\bibfnamefont {C.M.} \bibnamefont {Will}},
  }\href {http://cds.cern.ch/record/1639582} {\emph {\bibinfo {title}
  {{Gravity: Newtonian, post-Newtonian, relativistic}}}} (\bibinfo  {publisher}
  {Cambridge University Press}, \bibinfo {address} {Cambridge}, \bibinfo {year}
  {2014})\BibitemShut {NoStop}%
\bibitem [{\citenamefont {Landau} and \citenamefont
  {Lifschits}(1975)}]{Landau:1982dva}%
  \BibitemOpen
  \bibfield  {author} {\bibinfo {author} {\bibfnamefont {L.D.} \bibnamefont
  {Landau}} and \bibinfo {author} {\bibfnamefont {E.M.} \bibnamefont
  {Lifschits}}, }\href@noop {} {\emph {\bibinfo {title} {{The Classical Theory
  of Fields}}}}, \bibinfo {series} {Course of Theoretical Physics}, Vol.
  \bibinfo {volume} {Volume 2} (\bibinfo  {publisher} {Pergamon Press},
  \bibinfo {address} {Oxford}, \bibinfo {year} {1975})\BibitemShut {NoStop}%
\bibitem [{\citenamefont {Battaglieri} \emph
  {et~al.}(2017)}]{Battaglieri:2017aum}%
  \BibitemOpen
  \bibfield  {author} {\bibinfo {author} {\bibfnamefont {M.}~\bibnamefont
  {Battaglieri}} \emph {et~al.}, }\href@noop {} {  (\bibinfo {year} {2017})},
  \Eprint {http://arxiv.org/abs/1707.04591} {arXiv:1707.04591}\BibitemShut
  {NoStop}%
\bibitem [{\citenamefont {Blas} \emph {et~al.}(2017)\citenamefont {Blas},
  \citenamefont {Nacir}, and \citenamefont {Sibiryakov}}]{Blas:2016ddr}%
  \BibitemOpen
  \bibfield  {author} {\bibinfo {author} {\bibfnamefont {D.}~\bibnamefont
  {Blas}}, \bibinfo {author} {\bibfnamefont {D.L.} \bibnamefont {Nacir}},  and
  \bibinfo {author} {\bibfnamefont {S.}~\bibnamefont {Sibiryakov}}, }\href
  {\doibase 10.1103/PhysRevLett.118.261102} {\bibfield  {journal} {\bibinfo
  {journal} {\emph {Phys. Rev. Lett.}} }\textbf {\bibinfo {volume} {118}},
  \bibinfo {pages} {261102} (\bibinfo {year} {2017})}, \Eprint
  {http://arxiv.org/abs/1612.06789} {arXiv:1612.06789}\BibitemShut {NoStop}%
\bibitem [{\citenamefont {Bošković} \emph {et~al.}(2018)\citenamefont
  {Bošković}, \citenamefont {Duque}, \citenamefont {Ferreira}, \citenamefont
  {Miguel}, and \citenamefont {Cardoso}}]{Boskovic:2018rub}%
  \BibitemOpen
  \bibfield  {author} {\bibinfo {author} {\bibfnamefont {M.}~\bibnamefont
  {Bošković}}, \bibinfo {author} {\bibfnamefont {F.}~\bibnamefont {Duque}},
  \bibinfo {author} {\bibfnamefont {M.C.} \bibnamefont {Ferreira}}, \bibinfo
  {author} {\bibfnamefont {F.S.} \bibnamefont {Miguel}},  and \bibinfo {author}
  {\bibfnamefont {V.}~\bibnamefont {Cardoso}}, }\href {\doibase
  10.1103/PhysRevD.98.024037} {\bibfield  {journal} {\bibinfo  {journal} {\emph
  {Phys. Rev.}} }\textbf {\bibinfo {volume} {D98}}, \bibinfo {pages} {024037}
  (\bibinfo {year} {2018})}, \Eprint {http://arxiv.org/abs/1806.07331}
  {arXiv:1806.07331}\BibitemShut {NoStop}%
\bibitem [{\citenamefont {López~Nacir} and \citenamefont
  {Urban}(2018)}]{LopezNacir:2018epg}%
  \BibitemOpen
  \bibfield  {author} {\bibinfo {author} {\bibfnamefont {D.}~\bibnamefont
  {López~Nacir}} and \bibinfo {author} {\bibfnamefont {F.R.} \bibnamefont
  {Urban}}, }\href@noop {} {  (\bibinfo {year} {2018})}, \Eprint
  {http://arxiv.org/abs/1807.10491} {arXiv:1807.10491}\BibitemShut {NoStop}%
\bibitem [{\citenamefont {Maggiore}(2007)}]{Maggiore:1900zz}%
  \BibitemOpen
  \bibfield  {author} {\bibinfo {author} {\bibfnamefont {M.}~\bibnamefont
  {Maggiore}}, }\href {http://www.oup.com/uk/catalogue/?ci=9780198570745}
  {\emph {\bibinfo {title} {{Gravitational Waves. Vol. 1: Theory and
  Experiments}}}}, Oxford Master Series in Physics (\bibinfo  {publisher}
  {Oxford University Press}, \bibinfo {year} {2007})\BibitemShut {NoStop}%
\bibitem [{\citenamefont {Blanchet}(2014)}]{Blanchet:2013haa}%
  \BibitemOpen
  \bibfield  {author} {\bibinfo {author} {\bibfnamefont {L.}~\bibnamefont
  {Blanchet}}, }\href {\doibase 10.12942/lrr-2014-2} {\bibfield  {journal}
  {\bibinfo  {journal} {\emph {Living Rev. Rel.}} }\textbf {\bibinfo {volume}
  {17}}, \bibinfo {pages} {2} (\bibinfo {year} {2014})}, \Eprint
  {http://arxiv.org/abs/1310.1528} {arXiv:1310.1528}\BibitemShut {NoStop}%
\bibitem [{\citenamefont {Binney} and \citenamefont
  {Tremaine}(2011)}]{BinneyTremaine}%
  \BibitemOpen
  \bibfield  {author} {\bibinfo {author} {\bibfnamefont {J.}~\bibnamefont
  {Binney}} and \bibinfo {author} {\bibfnamefont {S.}~\bibnamefont {Tremaine}},
  }\href {https://books.google.ch/books?id=6mF4CKxlbLsC} {\emph {\bibinfo
  {title} {Galactic Dynamics: (Second Edition)}}}, Princeton Series in
  Astrophysics (\bibinfo  {publisher} {Princeton University Press}, \bibinfo
  {year} {2011})\BibitemShut {NoStop}%
\bibitem [{\citenamefont {Martinez} \emph {et~al.}(2017)\citenamefont
  {Martinez}, \citenamefont {Stovall}, \citenamefont {Freire}, \citenamefont
  {Deneva}, \citenamefont {Tauris}, \citenamefont {Ridolfi}, \citenamefont
  {Wex}, \citenamefont {Jenet}, \citenamefont {McLaughlin}, and \citenamefont
  {Bagchi}}]{Martinez:2017jbp}%
  \BibitemOpen
  \bibfield  {author} {\bibinfo {author} {\bibfnamefont {J.G.} \bibnamefont
  {Martinez}}, \bibinfo {author} {\bibfnamefont {K.}~\bibnamefont {Stovall}},
  \bibinfo {author} {\bibfnamefont {P.C.C.} \bibnamefont {Freire}}, \bibinfo
  {author} {\bibfnamefont {J.S.} \bibnamefont {Deneva}}, \bibinfo {author}
  {\bibfnamefont {T.M.} \bibnamefont {Tauris}}, \bibinfo {author}
  {\bibfnamefont {A.}~\bibnamefont {Ridolfi}}, \bibinfo {author} {\bibfnamefont
  {N.}~\bibnamefont {Wex}}, \bibinfo {author} {\bibfnamefont {F.A.}
  \bibnamefont {Jenet}}, \bibinfo {author} {\bibfnamefont {M.A.} \bibnamefont
  {McLaughlin}},  and \bibinfo {author} {\bibfnamefont {M.}~\bibnamefont
  {Bagchi}}, }\href {\doibase 10.3847/2041-8213/aa9d87} {\bibfield  {journal}
  {\bibinfo  {journal} {\emph {Astrophys. J.}} }\textbf {\bibinfo {volume}
  {851}}, \bibinfo {pages} {L29} (\bibinfo {year} {2017})}, \Eprint
  {http://arxiv.org/abs/1711.09804} {arXiv:1711.09804}\BibitemShut {NoStop}%
\bibitem [{\citenamefont {Stovall} \emph {et~al.}(2018)}]{Stovall:2018ouw}%
  \BibitemOpen
  \bibfield  {author} {\bibinfo {author} {\bibfnamefont {K.}~\bibnamefont
  {Stovall}} \emph {et~al.}, }\href {\doibase 10.3847/2041-8213/aaad06}
  {\bibfield  {journal} {\bibinfo  {journal} {\emph {Astrophys. J.}} }\textbf
  {\bibinfo {volume} {854}}, \bibinfo {pages} {L22} (\bibinfo {year} {2018})},
  \Eprint {http://arxiv.org/abs/1802.01707} {arXiv:1802.01707}\BibitemShut
  {NoStop}%
\bibitem [{\citenamefont {J.~Hailey} \emph {et~al.}(2018)\citenamefont
  {J.~Hailey}, \citenamefont {Mori}, \citenamefont {E.~Bauer}, \citenamefont
  {E.~Berkowitz}, \citenamefont {Hong}, and \citenamefont
  {J.~Hord}}]{2018Natur.556...70H}%
  \BibitemOpen
  \bibfield  {author} {\bibinfo {author} {\bibfnamefont {C.}~\bibnamefont
  {J.~Hailey}}, \bibinfo {author} {\bibfnamefont {K.}~\bibnamefont {Mori}},
  \bibinfo {author} {\bibfnamefont {F.}~\bibnamefont {E.~Bauer}}, \bibinfo
  {author} {\bibfnamefont {M.}~\bibnamefont {E.~Berkowitz}}, \bibinfo {author}
  {\bibfnamefont {J.}~\bibnamefont {Hong}},  and \bibinfo {author}
  {\bibfnamefont {B.}~\bibnamefont {J.~Hord}}, }\bibfield  {booktitle} {\emph
  {\bibinfo {booktitle} {Nature}}, }\href@noop {} { \textbf {\bibinfo {volume}
  {556}}, \bibinfo {pages} {70} (\bibinfo {year} {2018})}\BibitemShut {NoStop}%
\end{thebibliography}%

\end{document}